\documentclass[journal]{IEEEtran}

\usepackage{amsmath,amssymb}
\usepackage{soul}
\usepackage{url,subfigure,graphicx,wrapfig}
\usepackage{fixmath}
\usepackage{blindtext}
\usepackage{bm}
\usepackage[ruled,linesnumbered]{algorithm2e}
\usepackage{multirow}
\usepackage {xcolor}
\usepackage{pdfrender}
\usepackage{diagbox}
\usepackage{amsbsy}
\usepackage{booktabs}
\usepackage{threeparttable}

\newcommand*{\boldcheckmark}{%
  \textpdfrender{
    TextRenderingMode=FillStroke,
    LineWidth=.5pt, 
  }{\checkmark}%
}

\begin{document}
%
\title{Eve Said Yes: AirBone Authentication for Head-Wearable Smart Voice Assistant
\author{
\IEEEauthorblockN{Chenpei Huang, Hui Zhong, Pavana Prakash, Dian Shi, Xu Yuan, and Miao Pan}}
\thanks{C. Huang, H. Zhong, P. Prakash, D. Shi and M. Pan are with the Department of Electrical and Computer Engineering, University of Houston, Houston,
TX, 77204 USA. (e-mail: chuang30@uh.edu; hzhong5@uh.edu; pprakash3@uh.edu; dshi3@uh.edu; mpan2@uh.edu)}
\thanks{X. Yuan is with the Department of Computer and Information Science at the University of Delaware, Newark, DE 19716 USA. (e-mail: xyuan@udel.edu)}
}


\maketitle

\begin{abstract}
Recent advances in machine learning and natural language processing have fostered the enormous prosperity of smart voice assistants and their services, e.g., Alexa, Google Home, Siri, etc.  
However, voice spoofing attacks are deemed to be one of the major challenges of voice control security, and never stop evolving such as deep-learning-based voice conversion and speech synthesis techniques. 
To solve this problem outside the acoustic domain, we focus on head-wearable devices, such as earbuds and virtual reality (VR) headsets, which are feasible to continuously monitor the bone-conducted voice in the vibration domain.
Specifically, we identify that air and bone conduction (AC/BC) from the same vocalization are coupled (or concurrent) and user-level unique, which makes them suitable behavior and biometric factors for multi-factor authentication (MFA).
The legitimate user can defeat acoustic domain and even cross-domain spoofing samples with the proposed two-stage AirBone authentication. The first stage answers \textit{whether air and bone conduction utterances are time domain consistent (TC)} and the second stage runs \textit{bone conduction speaker recognition (BC-SR)}. The security level is hence increased for two reasons: (1) current acoustic attacks on smart voice assistants cannot affect bone conduction, which is in the vibration domain; (2) even for advanced cross-domain attacks, the unique bone conduction features can detect adversary's impersonation and machine-induced vibration.
Finally, AirBone authentication has good usability (the same level as voice authentication) compared with traditional MFA and those specially designed to enhance smart voice security.
Our experimental results show that the proposed AirBone authentication is usable and secure, and can be easily equipped by commercial off-the-shelf head wearables with good user experience.

\end{abstract}

%

\section{Introduction \label{sec:1_intro}}
With recent advances in human and natural language processing, smart voice assistants (VA) are widely equipped by various platforms, such as smartphones, computers, earbuds, and smart home devices. 
The ever-developing voice intelligence can understand and smoothly respond to the user's spoken commands, bringing tremendous convenience and reshaping daily lives. For instance, ``Call my friend'' or ``Play my favorite music'' can be recognized by Alexa, Google Home and Apple Siri, and those VAs can automatically execute the voice commands.
Furthermore, voiceprint authentication is widely accepted in many mobile and wearable scenarios. For example, ``Log me into the metaverse'' by metaverse systems is now readily supported by most metaverse devices. Many industry experts even predict that nearly every application will integrate smart voice control technology in some way in the next decade~\cite{VoiceControlEverything}.

\begin{figure}[t]
    \centering
    \includegraphics[width=0.48\textwidth]{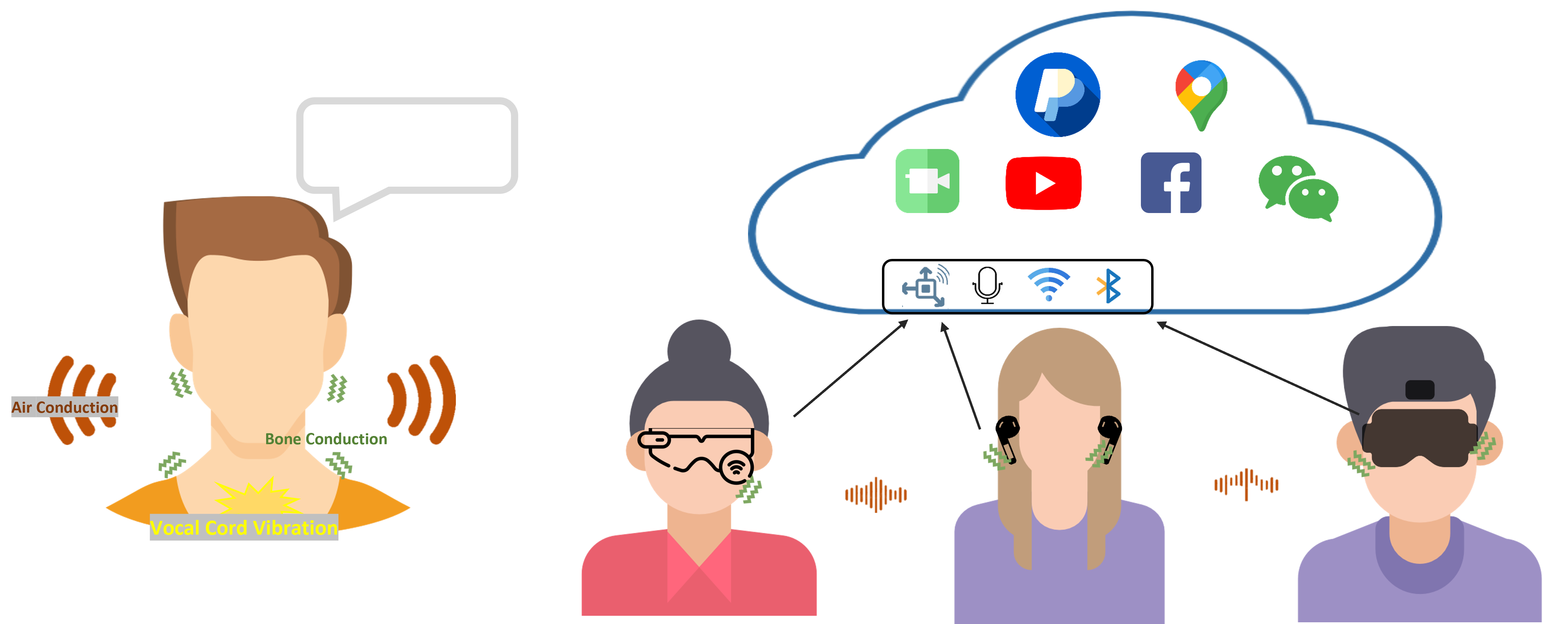}
    \caption{: (Left) Illustration of air and bone conduction signals in the human vocalization process. When the vocal cord vibrates, the air-conducted voice propagates over the air, while the bone-conducted voice causes speech-related skull/skin/soft tissue vibrations. (Right) Potential Usage. Users can call smart voice services by their head wearable/contact devices, e.g., smart glasses, earbuds, and virtual reality headsets. Bone conduction signals recorded by such devices will enhance voice control security.}
    \vspace{-5mm}
    \label{fig:1_teaser}
\end{figure}

However, the ideal ``voice control everything'' and ``voice recognize everyone'' are hindered by attacks targeting smart voice assistants. Those attacks have gradually been recognized as a research challenge by the cybersecurity community in the last decades~\cite{kamble2020advances,lau2004vulnerability,ergunay2015vulnerability,kinnunen2012vulnerability}. Specifically, \textbf{voice spoofing attacks} can generate malicious voice commands with spectral patterns very close to the legitimate user. Those spoofing methods include \textit{synthetic speech (SS)}, \textit{voice conversion (VC)}, \textit{replay}, and \textit{impersonation}. On the other hand, \textbf{audio injection attacks} mix the genuine speaker's voice commands with \textit{hidden voice commands} or \textit{audio adversarial patterns} to mislead the deep-learning-based automatic speech recognition or speaker recognition models (ASR/SR) towards wrong decisions. Tremendous research efforts have been made to shield smart voice assistants from the aforementioned attacks, for example, by accurate and robust speaker recognition. Unfortunately, it tends to become an ``arms race'' because new attacks also leverage powerful machine-learning technologies (e.g., generative artificial intelligence, GenAI). As a result, knowing the vulnerability of smart voice assistants, the attacker imposes threats to the victim's life and property safety. Moreover, potential customers may be reluctant to disclose information to the voice assistant and unable to make full use of it.

Recently, another line of research has tended to protect VA beyond the acoustic domain. One straightforward solution is to bind with biometrics (e.g., fingerprint, vein, iris, etc.) or non-biometrics (e.g., password, hardware token, one-time code, etc.) factors, known as multi-factor authentication (MFA) \cite{multifactor,manzoor2019multi}, e.g., Duo Mobile \cite{DuoMobile}. 
However, traditional MFA goes against the original intention of voice interaction, which enables immersive and smooth human-computer interaction. The extra efforts and time in extra data acquisition from users may significantly downgrade user experiences and discourage customers from using those systems. To tackle this issue, researchers found that voice-induced signals/phenomena, which are concurrently generated during the human vocalization process, potentially improve and secure acoustic-domain voice authentication. Briefly, those factors answer ``the user \textit{did} say it.'' To name a few, VAuth \cite{VAuth} by Feng et al. and WearID \cite{shi2020wearid} by Shi et al. focused on the cross-correlation between voice and voice-induced vibration collected from the wearable's onboard accelerometer. Also, Earprint \cite{gao2021voice} proposed by Gao et al. utilized in-ear voice as biometrics. Nevertheless, these pioneer works still have their limitations. Most importantly, they assume the attacker cannot obtain the cross-domain factors, and therefore cannot launch cross-domain attacks. In this paper, we choose to select bone conduction as the second domain because of its accessibility on head-wearable devices, concurrency with (air-conducted) speech, and user-level uniqueness against cross-domain attacks.

We now revisit the human vocalization process. There are several simple but intriguing observations. When people speak, their vocal cords generate vibrations. As the vibrations are carried through the air, the speech signals pass through filters formed by the mouth, tongue, and nose. These signals can propagate over the air and be heard by human listeners and microphones, known as air conduction signals (AC). On the other hand, as the vibrations propagate inside the human body (i.e., bones, soft tissues, etc.), these signals can be received by attached motion sensors (e.g., accelerometers and gyroscopes), known as bone conduction (BC) signals \cite{liu2018vocal,schneegass2016skullconduct,BC2000}. For head-wearable devices, microphones and motion sensors are currently available on most commercial off-the-shelf platforms. In other words, both air and bone conduction (air-bone) signals can be successfully collected by the head-wearable devices illustrated in Fig.~\ref{fig:1_teaser}. The reasons we choose the air and bone conduction domain for cross-domain authentication are because (1) BC can only be coupled with spoken AC (rather than environmental noise); (2) BC contains speech and phoneme information (compared with password or fingerprint); 3) BC signals are also unique biometric because of
different attenuation properties among different users, incurred by
different skull structures, bone density, and so on \cite{BC2000}.

Inspired by BC signals' salient features above, in this paper, we propose a novel AirBone cross-domain authentication design for smart VA on head-wearable devices. Briefly, we equip the users with earbuds, smart glasses, and virtual reality (VR) headsets\footnote{The motion sensors have already been incorporated into advanced earbuds/headsets, though such designs are not for authentication purposes, e.g., Apple AirPods has integrated the high-end motion sensor to improve voice call quality in noisy environments.} all having accelerometers. 
We first develop an air-bone temporal consistency scoring (TCS) algorithm, which helps VA identify whether the air-bone domain signals are from the same user’s vocalization. Next, we develop a learning-based BC speaker recognition (BC-SR), which can distinguish legitimate users from spoofing ones in the BC domain. Finally, based on our self-collected dataset, we evaluate the proposed AirBone authentication design under multiple adversarial scenarios. The empirical results suggest that the proposed AirBone authentication is usable and secure in practical settings and outperforms the recent peer designs in many aspects. 

Overall, our salient contributions are summarized as follows.
(1) We leverage BC signals to develop a novel secure and “effortless-to-use” cross-domain authentication specified for VA on head wearable devices. Our method exploits multiple useful authentication factors from the air-bone signals and only requires users to speak for authentication.
(2) We propose a two-stage authentication TCS (stage I) and BC-SR (stage II), to verify the microphone received voice command: 1) is spoken by the user operating the head wearable device (not an acoustic domain imposter), 2) is spoken by the legitimate user (not a vibration domain imposter). We develop the TCS algorithm to measure the temporal consistency between two domains and distinguish the attack from benign samples. Once the tested air-bone pairs pass the first stage, the BC-SR can verify the legitimate user’s BC spectral patterns, which are much harder to spoof compared with AC signals.
(3) We conduct extensive experiments to validate the effectiveness and evaluate the performance of the proposed authentication design under multiple adversarial scenarios, including noises and spoofing attacks. The experimental results demonstrate the proposed air-bone authentication is secure and surpasses the peer designs in the usability aspect.

The rest of the paper is organized as follows. Section.~\ref{sec:2_background} presents the background of cross-domain authentication for VA and bone conduction. Section~\ref{sec:3_adversarial} and \ref{sec:4_efficacy} introduce the adversarial and the efficacy of the proposed design. In Section~\ref{sec:5_design}, we detail the proposed authentication method in two stages, namely TCS and BC-SR. The performance and robustness are evaluated in Section~\ref{sec:6_eval}, with the discussion in Section~\ref{sec:7_discuss} and the conclusion in Section~\ref{sec:8_conclusion}.

\section{Background\label{sec:2_background}}
\subsection{Cross-Domain Authentication}
Smart voice services can authorize users via their spoken speech, known as speaker verification. However, since the air-conducted speech (AC) signal propagates in the open medium (namely, air), the attacker's spoofing samples can be received by the VA's microphone. Many existing studies have discussed the attacks and defenses in the acoustic domain \cite{kinnunen2012vulnerability,ergunay2015vulnerability}. Among them, one promising solution is to investigate concurrent signals during vocalization. 

Feng et al. proposed VAuth in \cite{VAuth}, which successfully utilized face vibrations matched with the voice to authenticate voice commands. However, segment-level analysis requires sufficient long cross-domain inputs to prevent performance degradation. WearID proposed by Shi et al. in \cite{shi2020wearid} selected air-induced smartwatch vibration for matching. Although no training is needed, such system usability highly relies on good acoustic and vibration signal strengths. Both of these methods cannot provide biometric factors, which are vulnerable to spoofing attacks \cite{wu2015asvspoof,wang2020asvspoof,gong2019remasc}. In fact, an adversary can launch a simple yet effective cross-domain attack to bypass the matching-based authentication, see Section.~\ref{sec:3_adversarial}. On the other hand, Gao et al. proposed Earprint in \cite{gao2021voice} by employing in-ear voice as a biometric. However, the selected two-domain signals are not well insulated, as the useful in-ear voice for authentication might be mixed with the audio signal played by other applications. As a result, such authentication cannot be performed when the earphones are playing music or are on phone calls. Zhang et al. in \cite{zhang2017hearing} leveraged the Doppler effect generated from the jaw and tongue movement in speaking. It required users to hold and move the mobile phone near their mouth until a perfect position. Such an approach can be promising for user authentication and liveness detection, but again, necessitating a large amount of user effort in data collection should always be avoided.

\subsection{Bone Conduction in Vibration Domain}
The bone conduction signal, also known as bone-borne vibration and speech-related facial vibration, originates from acoustic wave propagation in solid material. Briefly, during vocalization, while voice is produced by vocal cord vibrations, filtered, and finally propagates over the air, the energy from the same vibrations also propagates along solid materials inside the body. The fundamental equation governing this phenomenon is the Navier{'}s equation, shown as
\begin{equation}
    \mu \nabla^2 \textbf{u} + (\lambda +\mu)\nabla(\nabla \cdot \textbf{u}) = \rho \left( \frac {\partial ^2 \textbf{u}} {\partial t^2} \right),
\label{eq:wave equation}
\end{equation}
where $\textbf{u}$ is the displacement vector, determined by solid material density $\rho$, Lam\'{e} constants $\lambda$ and $\mu$, and structure boundary conditions. 

\begin{figure}[t]
\centering
    \subfigure[AC voice spectrogram.\label{fig:2_ac_spec}]
    {\includegraphics[width=.48\linewidth]{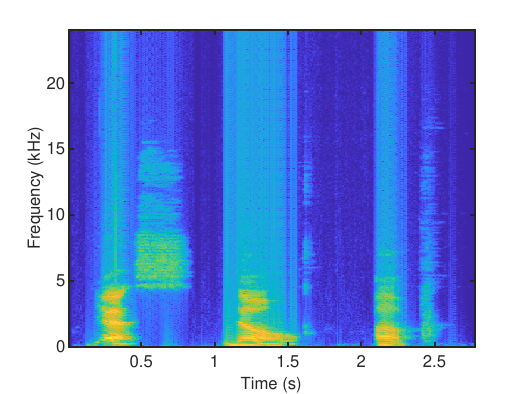}}
    \subfigure[BC voice spectrogram.\label{fig:2_bc_spec}]
    {\includegraphics[width=.48\linewidth]{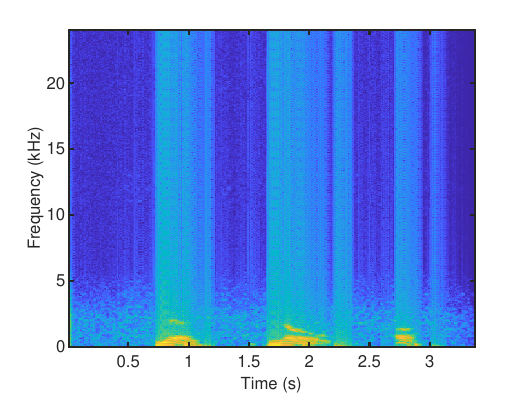}}
    \caption{The spectrograms of a pair of air-bone signals, containing the same voice command “one, two, three” (not synchronized). Note that the BC signal almost has no components above 2kHz, whose frequency range is narrower than the AC signal’s}
    \label{fig:2_spec}
\end{figure}

Equation~(\ref{eq:wave equation}) has demonstrated the close relationships between the AC/BC signals. However, the discrepancy in propagation media eventually makes AC and BC sound differently and categorizes them as two domains. As stated in \cite{BC2010}, BC signals encounter frequency and phoneme selectivity due to position when mouth closing and opening. In recent papers, \cite{BCEnhance} and \cite{BCSurvey} model the AC/BC signals as two different versions of filtered ``clear signal'', shown as
\begin{equation}
\begin{aligned}
    y_a(f,t) &= x(f,t) + n(f,t), \\ 
    y_b(f,t) &= g(x(f,t)) + w(f,t),
\end{aligned}
\label{eq:bc_model}
\end{equation}
where $y_a$ and $y_b$ indicate the AC and BC signals, respectively, $x$ is the clear signal, $n$ and $w$ are AC and BC noises, $f$ and $t$ represent the frequency and time index of the Short-Time Fourier Transform (STFT). The non-linear function $g$ in this model describes the discrepancy between AC and BC signals and makes them two separate domains. We plot the AC and BC spectrogram from the sample voice command ''one, two, three'' in Fig.~\ref{fig:2_spec}, where the brighter color represents larger spectral energy. We also notice that, unlike AC (audio) signals covering a wider spectrum in Fig.~\ref{fig:2_ac_spec}, the BC signal only dominates the spectrum below 2 kHz in Fig.~\ref{fig:2_bc_spec} from our self-collected datasets, and presents a very different time-frequency pattern compared with AC's.

\begin{figure*}[t]
        \centering
        \includegraphics[width = 0.89\linewidth]{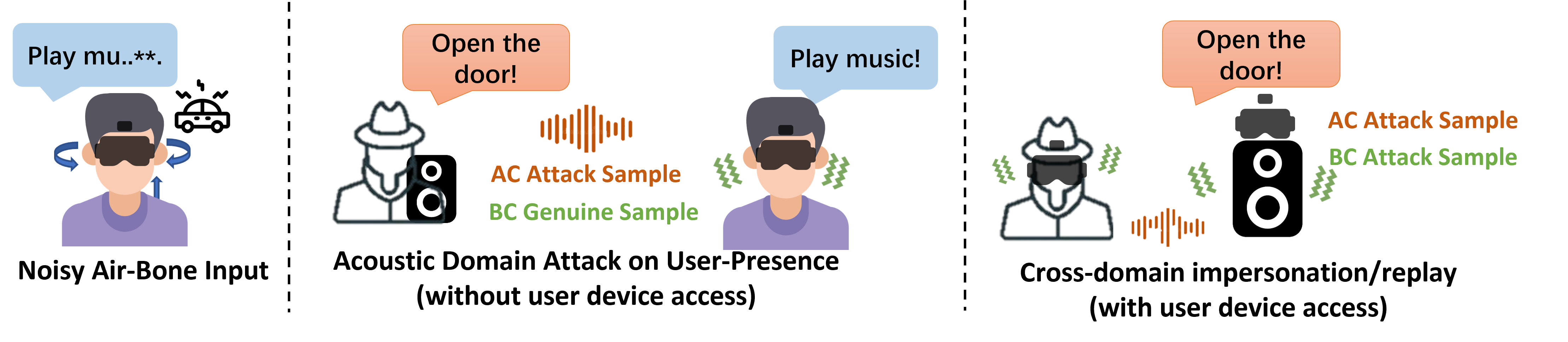}
        \caption{Illustration of three adversarial scenarios. Noisy air-bone input (left): the legitimate user uses voice commands in a noisy environment. Acoustic domain attacks (mid): the legitimate user owns the required wearable devices. Cross-domain attacks (right): the attacker gets access to the required wearable devices.}
        \label{fig:4_threat_model}
\end{figure*}

\section{Adversarial Scenarios\label{sec:3_adversarial}}

The adversarial scenarios are illustrated in Fig.~\ref{fig:4_threat_model}, namely, \textbf{noisy air-bone input}, \textbf{user-present acoustic domain attack},\textbf{ and user-absent cross domain attack}. We assume the hardware, including the sensor and the computing/memory unit, are inviolable and resistant to tampering. We also assume the sanctity of any possible wireless communication during AirBone authentication. This assumption is valid thanks to the secure wireless transmission protocols, such as Wi-Fi Protected Access 3 \cite{WPA3}, Bluetooth Service Level 4 \cite{Bluetooth}, Kerberos \cite{kerberos}, Transport Layer Security (TLS) \cite{TLS}, and so on. Therefore, the data packet will not be intercepted or manipulated by the attacker. In addition, the history sensory signals will not be stored in the local or remote device, thereby precluding potential risks associated with unauthorized data access or breaches.

We also make several assumptions about the attacker's knowledge and capability. First, the attacker's target is to defeat the voice-based authentication systems so that he can access personal information and virtual assets. Second, we assume the attacker will not physically compromise the system. Third, we contend that the attacker can initiate attacks on the user's presence and absence. When the user is present (also, the wearable device in use), the attacker can only launch stealthy acoustic domain attacks to avoid notifying the user. When the user is absent (wearable device controlled by the attacker), the attacker leverages insights gained from prior observation or the device user manual to log in.

\smallskip
\noindent\textbf{Air-Bone Domain Noise.}
Due to air being an open medium, the microphone will record various speech-unrelated signals from the environment. Besides, speech-related signals from background speakers can also exist, such as background speakers, TV sets, and radios. We can use the following equation to model the signal and noise in the AC domain:
\begin{equation}
\label{eq:ac_noise}
    r_{noisy} = y_a + \sum_{i=1}^{N}b_{i} + n,
\end{equation}
where $r_{noisy}$ denotes the noisy microphone recordings, $y_a$ denotes the desired AC signal from the user, $b_i$ denotes the $i$-th background user, and $n$ denotes the environmental noise. In practice, plenty of denoising methods can be applied to the raw AC recordings for stationary noise removal, for instance, spectral subtraction and Wiener Filtering. Furthermore, mitigating the impact from background users falls into the area of speaker diarization and audio source separation. Therefore, we can expect a small performance degradation caused by AC domain noise when employing all the techniques above.

BC domain noise ($w$ in Eq.~(\ref{eq:bc_model})) includes bone-conducted body motions, gestures, gravity and thermal noise from Micro-electromechanical Systems (MEMS) accelerometer. We employed signal preprocessing, noise-resilient TCS in Stage I and adversarial learning for Stage II authentication to mitigate the negative impacts from BC domain noise.

\smallskip
\noindent\textbf{User-Present: Acoustic Domain Attack.}
The first targeted attack model also follows as the acoustic domain attack. As discussed in the literature \cite{kamble2020advances,wu2015asvspoof,wang2020asvspoof}, four spoofing attacks are summarized as follows. (1) \textit{Impersonation attack.} The adversary produces a similar voice pattern and speech behavior as the target speaker. (2) \textit{Synthetic Speech (SS) attack.} The adversary inputs texts and generates corresponding human voices using the computer. (3) \textit{Voice Conversion (VC) attack.} The adversary converts his/her own voice to a sound similar to the victim speaker. (4) \textit{Replay attack.} The adversary records the legitimate user's voice samples and replays them for access. To summarize, the goal of spoofing attacks is to mimic the legitimate user's voice as much as possible. In this paper, we select the impersonation and replay attacks to represent acoustic domain attacks.

\smallskip
\noindent\textbf{User-Absent: Cross Domain Attack.}
Although it is often assumed that the attacker can not generate consistent cross-domain signals \cite{VAuth,shi2020wearid}, we find the attacker can easily do so when the legitimate user is absent. The acoustic domain spoofing attack can be easily extended to cross-domain ones as follows. (1) \textit{Air-bone cross-domain impersonation attack.} The adversary puts on wearable devices to launch impersonation attacks. (2) \textit{Air-bone cross-domain SS/VC/replay attack.} The adversary puts the wearable devices very close to or even on the surface of the synthetic voice generator/voice converter/recorder to launch an SS/VC/replay attack, which is temporal consistent in both domains.

\begin{figure*}[t]
    \centering
    \includegraphics[width=0.7\linewidth]{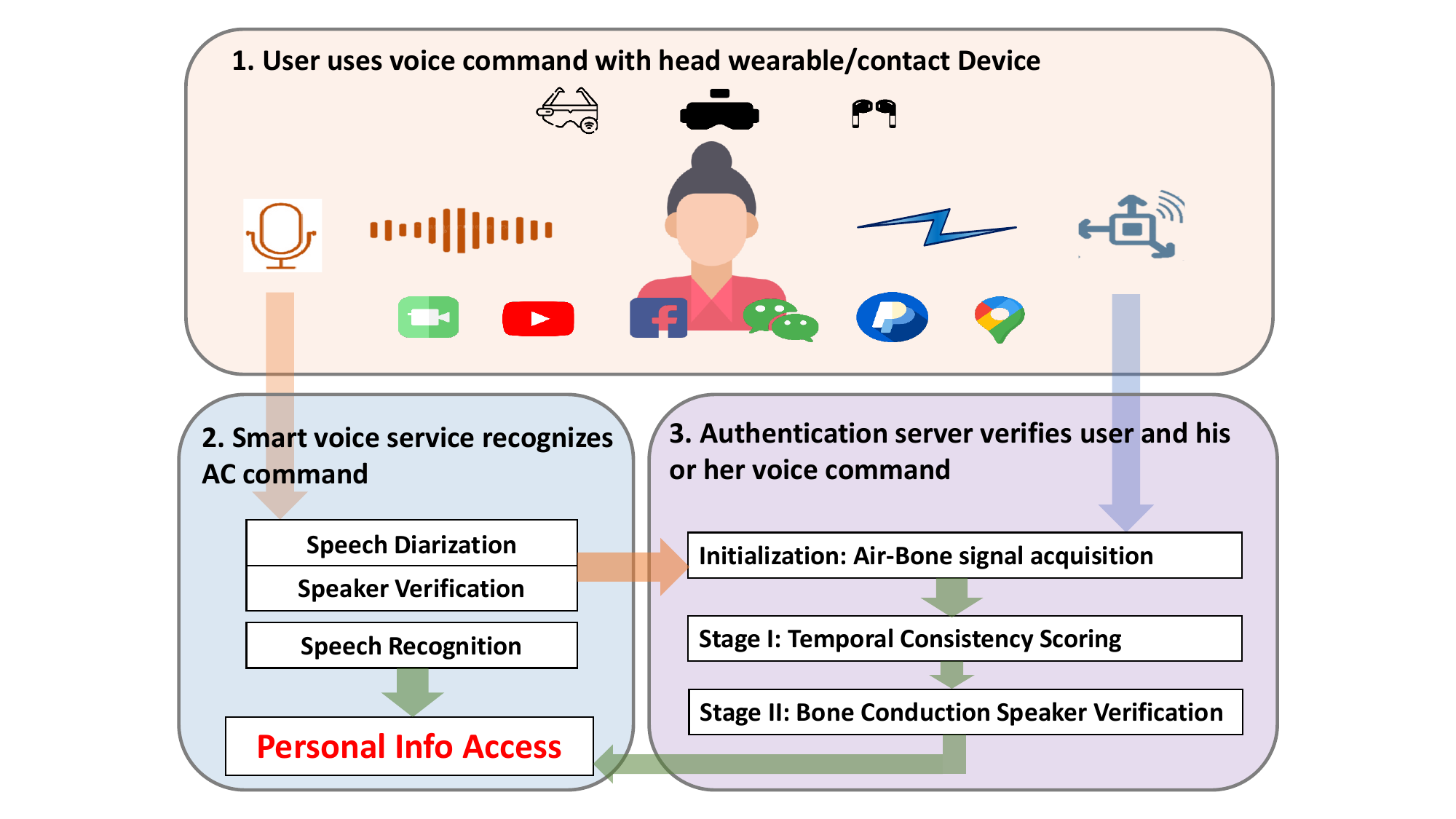}
    \caption{Design overview. Step 1: The user initiates the smart voice services using his/her voice commands. Step 2: The service provider recognizes the voice commands and sends an authentication request to the authentication server. Step 3: The authentication server gathers the air-bone signals and then runs a two-stage authentication to enhance the voice command security.}
    \vspace{-5mm}
    \label{fig:3_design_overview}
\end{figure*}

\section{Efficacy of the Proposed Approach\label{sec:4_efficacy}}
\subsection{Bone Conduction Security Features}
The bone conduction voice provides a number of appealing features for security applications. \textbf{\textit{Non-Propagating.}} The BC signals are essentially acoustic reverberation in a solid medium (human bone and soft tissues). Due to the acoustic impedance greatly diverging in bone and air, the BC signals will not propagate over the air and mix with the air-conducted ones. Meanwhile, the AC signals, especially mixed with multiple noise sources, will have no impact on BC signals. Therefore, the denoising process can be separately done in each domain, while more importantly, the BC signals will remain secure under acoustic spoofing and injection attacks. \textbf{\textit{Physical Contact.}} The BC signals are extracted from tiny facial vibrations during the vocalization process, which requires contact microphones or motion sensors for signal acquisition. In other words, it can be almost impossible for an attacker to record the genuine BC signals without notifying the legitimate user\footnote{The attacker may choose to use wireless sensing methods, e.g., mmWave, to detect the BC signals from tiny vibrations. However, mmWave only works well with line-of-sight (LOS), which can be easily perceived by the legitimate user}. \textbf{\textit{Structure-Sensitive.}} In previous literature \cite{BC2000,BC2010}, BC is modeled as a non-linear channel determined by bone and soft tissue structures. Each user has his or her own BC channel, and based on this, an effective user identification framework was proposed \cite{schneegass2016skullconduct}. In our design, we claim the structure-sensitive characteristics are able to detect BC impersonation and machine-induced vibration when the attacker gains physical access to the target device and wears it on. 

\subsection{Core Idea}
The proposed AirBone authentication consists of \textit{\textbf{Initialization}} to preprocess the raw data samples and two authentication stages: \textit{\textbf{Temporal Consistency Scoring (Stage I: TCS)}} to verify whether the air-bone data are from the same vocalization process; \textit{\textbf{Bone-Conduction Speaker Recognition (Stage II: BC-SR)}} to verify whether the speaker is the legitimate user. It is worth noting that we do not discuss air-conduction speaker verification (AC-SV) since it has been well-studied and already adopted by the current VA. Instead, our proposed method is to enhance VA security.

\subsection{Challenge}
\begin{itemize}
    \item The raw acceleration data is dominated by gravity, body motions, and noises, which can make it hard to extract useful BC signals. 
    \item The short and noisy AC and BC signals make it challenging to evaluate the consistency between the two domains as a behavioral factor.
    \item Cross-domain attack can defeat previous matching-based authentication, e.g., VAuth\cite{VAuth}. This is because when accessing the device, the attacker's own/converted voice is consistent with his own BC voice.
    \item As a result, the bone-conduction uniqueness should be included in the authentication process. However, no current BC-SR methods can succeed considering normal use and attacks mentioned in Section.~\ref{sec:3_adversarial}.
\end{itemize}

\section{AirBone Authentication Design\label{sec:5_design}}
In this section, we will first introduce the sensors used to capture AC and BC signals and present how to extract BC signals from raw motion data, namely the Initialization stage. Next, we introduce the two-stage algorithm to authenticate a user. At Stage I, we evaluate the air-bone signal consistency at certain representative frequencies and propose a time-domain consistency scoring (TCS) algorithm. At Stage II, we exploit effective bone conduction user recognition with pretraining and domain-adversarial learning.

\subsection{Initialization: Air-Bone Signal Acquisition \label{sec:air-bone design:s0}}
\begin{figure}[t]
        \centering
        \includegraphics[width = 0.9\linewidth]{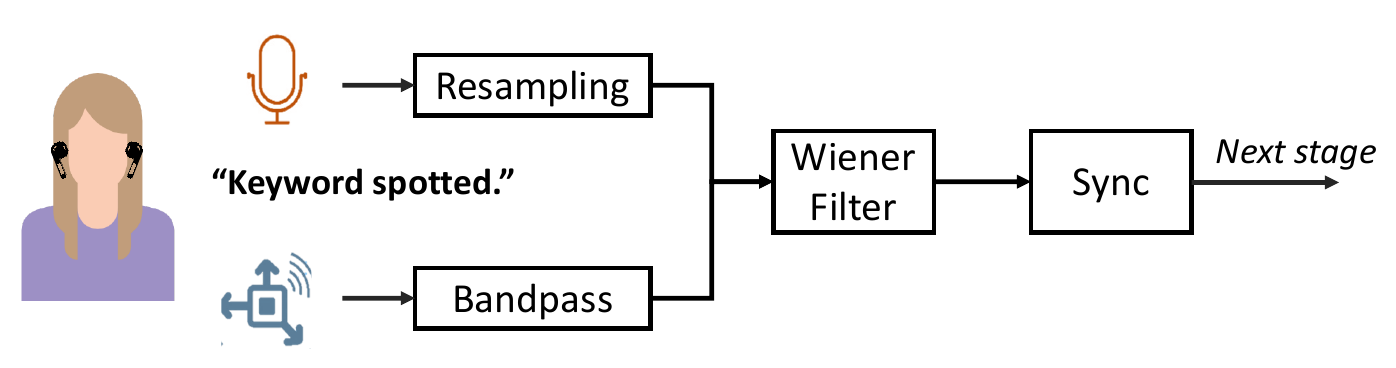}
        \caption{Flow chart of Initialization. The air-bone signals are recorded once the system detects a preset keyword, e.g. ``Okay Google''. The air conduction signals will be resampled to 8kHz to match the sampling frequency of BC signals. Next, after Wiener Filtering, the denoised air-bone signals will be synchronized using frame-level cross-correlation.}
        \label{fig:5_init_flowchart}
    \end{figure}

The flowchart of Initialization is shown in Fig.~\ref{fig:5_init_flowchart}. We assume the microphone and the accelerometer, as the receivers of air-bone signals, can wake up by the keyword spotting at the same time. The goal of initialization is to preprocess the raw input recordings into clear and synchronized AC and BC signals.

\smallskip
\noindent\textbf{AC Signal Processing.} When speaking, the user's voice will propagate as acoustic waves and will be recorded by microphones on head wearable devices. Although different types of microphones differ in components and sizes, all of them can convert incoming acoustic waves to electrical signals. For example, by using a diaphragm as a flexible capacitor plate, the changing current reflects the diaphragm vibration and hence records the sound wave. Since the electrical signal generated in this method is analog, it requires a preamplifier, a codec with an analog-to-digital converter (ADC), and a digital signal processor (DSP). One typical sampling frequency is 48 kHz. We downsample the AC signals to an 8 kHz sampling rate to reduce the data size and make it consistent with the BC signal sampling rate. Compared with the ``clear'' AC signal in Eq.~(\ref{eq:bc_model}), the AC recording should consider the impulse response determined by distance and orientation between the user and the recording device. 

\smallskip
\noindent\textbf{BC Signal Processing.} To record the BC signals, the required sensor must be capable of contacting the human body and sensing the tiny vibrations caused by human speaking. One common solution is to employ a built-in accelerometer, which has been integrated into most commercial (head) wearable devices, such as earbuds, AR/VR headsets, and smart glasses. 
Although accelerometers in the above wearable devices are widely used for motion tracking or speech applications (e.g., speech detection in Airpods), they have seldom been used for authentication purposes.
In this paper, the high-resolution and low-noise accelerometer LIS25BA is employed for evaluation. However, many other MEMS motion sensors (e.g., MPU-6500, MPU-9150, LSM6DSL) have the capacity to produce equivalent results, according to their technical specifications such as sensitivity, sample rate, and programmable filter parameters.

The BC signal requires additional signal processing to extract the speech-related vibration from the raw accelerometer readings. First, an accelerometer is a motion sensor, which outputs the measured acceleration from three directions. Such a problem is addressed by utilizing the direction perpendicular to the skin. In this work, we simply remove the gravity and slow body motions by a high-pass filter with a 20 Hz cut-off frequency. Second, the highest frequency of interest should be limited to 2 kHz, shown in Fig.~\ref{fig:2_bc_spec}. Third, to further increase the signal-to-noise ratio (SNR), the noise reduction technique can be applied to the acceleration signals. Specifically, we employ the adaptive Wiener filter to estimate noise from unvoiced segments and then suppress noise from voiced ones using the estimated statistics. Thus, the overall signal processing of the BC signal can be expressed as
\begin{equation}
\label{eq:BC_pre}
    y_{b}(t) = \text{Normalize}(\beta(t)) * h_{bp}(t) * h_{W}(t),
\end{equation}
where $y_b$ denotes the extracted bone conduction signal, $\beta$ represent the original accelerometer readings, $h_{bp}$ indicates the impulse response of the bandpass filter, and $h_W$ denotes the impulse response of Wiener filter. Symbol $*$ in the equation means time-domain convolution. 

\smallskip
\noindent{\textbf{Synchronization.}} As a prerequisite for evaluating air-bone domain signals' temporal consistency, we use cross-correlation for synchronization. Specifically, we select the first $K$ frames in succession from both AC and BC signals. Following the frame orders as $1, 2, \ldots, K$, the cross-correlation is calculated between the current AC and BC frames. Next, by locating the position $\tau_k$ yielding the maximum cross-correlation value between $k$-th AC/BC frames, we estimate the delay $\hat{\tau}$ by averaging $\tau_k$, shown as
\begin{equation}
    \hat{\tau} = \frac{1}{K} \sum_{k=1}^{K} \mathop{\arg\max}_{\tau} \left( \int_{t=-L/2}^{L/2} y_{a}(t) y_{b}(t-\tau) dt \right),
\end{equation}
where $\hat{\tau}$ denotes the estimate delay, $K$ denotes the number of frames, and $L$ denotes the frame length. Based on our empirical study, the estimation error vanishes when increasing the frame length and the number of frames. Finally, with the estimated delay, we can align the starting time precisely by padding zeros to the beginning of the earlier signal.
\subsection{Stage I: Temporal Consistency Scoring (TCS) \label{sec:air-bone design:s1}}
We evaluate the consistency of air-bone signals to verify if AC and BC signals are from the same speaker's vocalization at this stage. If the air-bone signals are from the same speaker, air-bone signals should have a strong temporal correlation after some transformations. It is intuitive to use 1-D (time) and 2-D (time-frequency) signal cross-correlation as mentioned in \cite{VAuth}, \cite{shi2020wearid}. However, the signal correlation between coupled air-bone signal pairs cannot always be distinguished from that of unrelated pairs. Recap that the BC signal generation process involves non-linear functions in Eq.~(\ref{eq:bc_model}), which generates several harmonic frequency components. When running a 1-D cross-correlation, as a simplified example, a pure tune in AC domain $y_a = \cos(2\pi ft)$ correlating with $y_b = (y_a)^2 = \frac{\cos(4\pi ft+1)}{2}$, will lead to 0, which can not be identified from the noise, regardless of the authenticity. In other words, the signal correlation is not stable to reflect air-bone signal consistency in the time domain. The success of VAuth \cite{VAuth} assumes a relatively long input length to ensure a sufficient number of ''surviving segments'', depending on the frequency of certain phonemes having strong cross-domain correlations. As a result, VAuth's accuracy will drop when encountering short voice commands, i.e., less than 6 seconds.

    \begin{algorithm}[t]
	\caption{Air-Bone Temporal Consistency Scoring (TCS) Algorithm.}
	\label{alg:algorithm1}
	\KwIn{\\Air conduction (AC) signal: $y_a(f,t)$; \\Bone conduction (BC) signal: $y_b(f,t)$;\\Parameter $ M, N$;}
	\KwOut{The consistency score $S$.}  
	\BlankLine
	Compute marginal BC power distribution w.r.t. time: $y_b(t)\leftarrow \sum_f y_b(f,t)$\;
	Remove silence at the beginning and end, $t'$ remains\;
	Compute marginal AC/BC power distribution w.r.t. frequency: $y_a(f)\leftarrow \sum_{t'} y_a(f,t')$, $y_b(f)\leftarrow \sum_{t'} y_b(f,t')$\;
	Select Top-M and Top-N frequency index: $m\leftarrow \arg_l \textit{Sort}\big(y_a(f)\big)_{1:M}$, $n\leftarrow \arg_l \textit{Sort}\big(y_b(f)\big)_{1:N}$\;
	Compute the correlation matrix: $C \leftarrow \textit{Corr}^{M\times N} \big[  y_a(f,t'),  y_b(f,t')                \big]$\; 
	Compute the consistency score: $S \leftarrow \max_{m \leq M, n \leq N} C(m,n)$.
\end{algorithm}
 
\smallskip
\noindent{\textbf{Insight.}} 
The insights of our proposed air-bone authentication solving the aforementioned problem are to focus on the spectral energy distribution in both air-bone domains. We first transform the time domain signals into time-frequency representations, e.g., short-time Fourier Transform (STFT). Therefore, the correlation between frequencies can be analyzed. Also, noticing the signal-to-noise ratio in each frequency bin is different, we select the representative frequencies and evaluate their statistical temporal correlation between the two domains to achieve a more reliable consistency score for authentication. Unlike 2D cross-correlation, the statistical temporal correlation among selected frequencies measures how closely these features behave in the time domain and save the computation on calculating the value traverse every time-frequency index. 

\smallskip
\noindent{\textbf{Scoring Algorithm.}} 
The proposed consistency scoring algorithm is shown in Algorithm~\ref{alg:algorithm1}. Synchronized in the Initialization, we normalize and frame both the air-bone signals, and then transform them by time-frequency analysis. We compute the distributions of AC/BC signal power w.r.t. the frequency. Both air-bone representative frequencies, which possess the most energy of the signal, are selected from air-bone signals, respectively. After that, we evaluate the statistical correlation among the coefficients of air-bone representative frequencies along the temporal axis. The maximum value in the resulting correlation matrix is described as the consistency score, which indicates the inherent relationship between air-bone signals. If air-bone signals are from the same speaker, the score will be high to indicate temporal consistency. Alternatively, if the air-bone signals are inconsistent, the score will be low to reject false triggering or attacks. 

   \begin{figure}[t]
   \centering
        \subfigure[STFT. \label{fig:6_bc_stft}]
        {\includegraphics[width=0.32\linewidth]{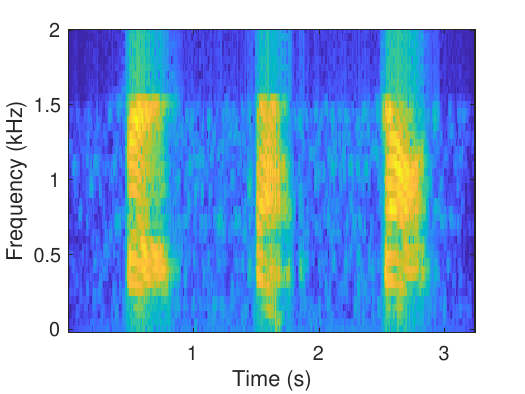}}
        \subfigure[CQT. \label{fig:6_bc_cqt}]
        {\includegraphics[width=0.32\linewidth]{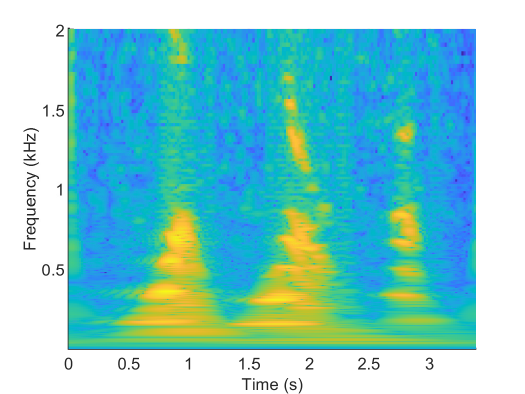}}
        \subfigure[x-axis reflection \label{fig:6_bc_cqt_aug}]
        {\includegraphics[width=0.32\linewidth]{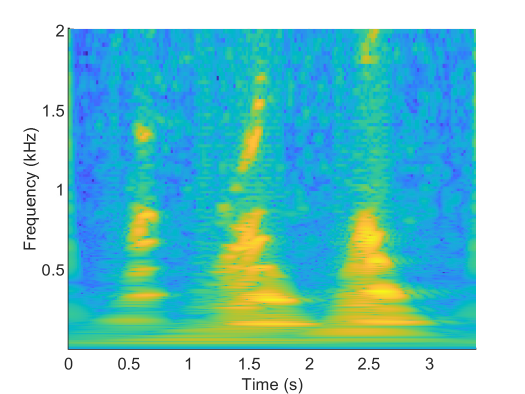}}
        \caption{Comparison between STFT, CQT, and CQT with data augmentation, as BC-SR input.}
        \label{fig:6_cqt_stft}
    \end{figure}

\begin{figure}[t]
    \centering
    \includegraphics[width = 0.99\linewidth]{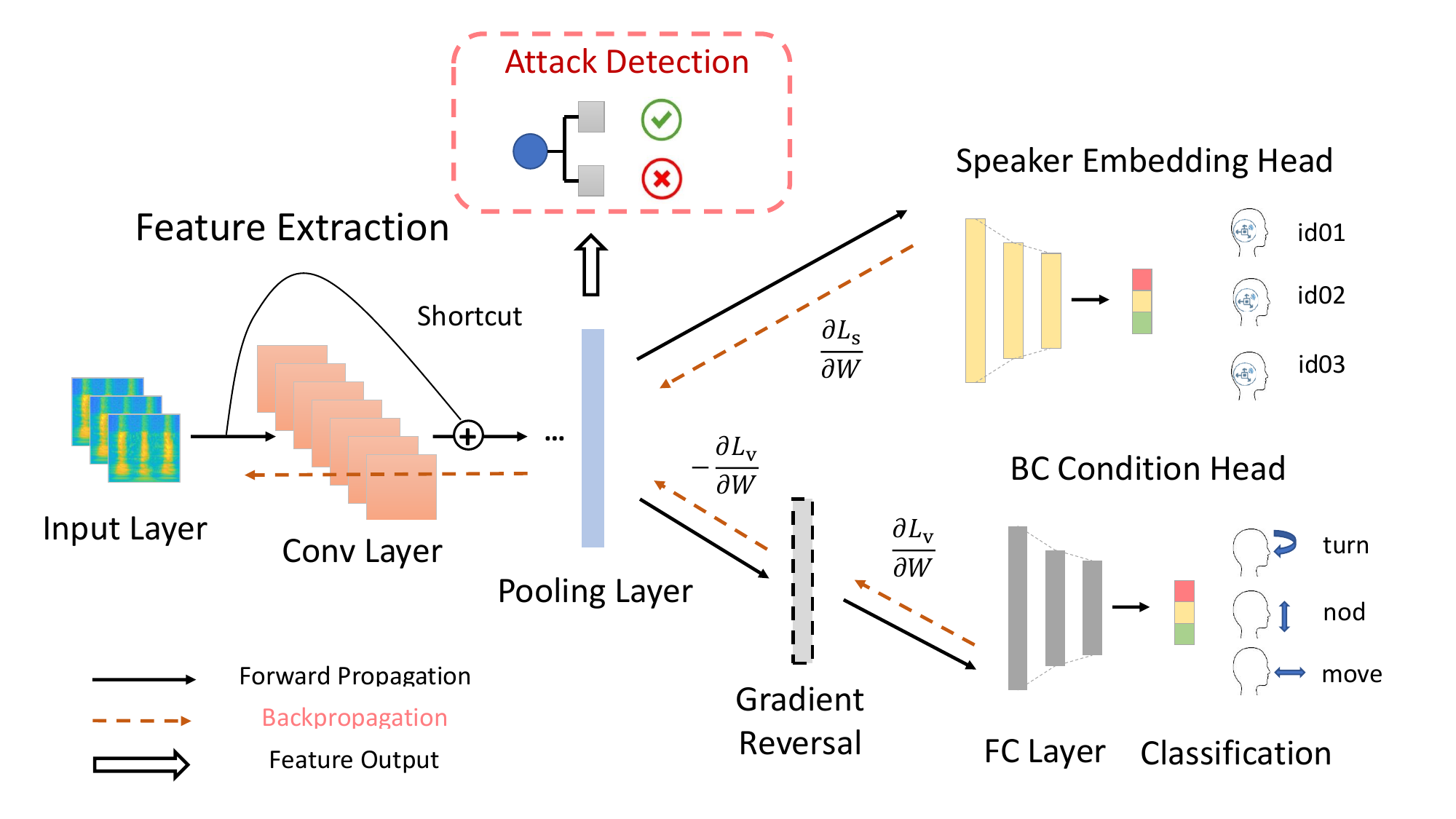}
    \caption{Illustration of the network structure. The convolutional layers serve as a feature extractor. The speaker embedding head trains the recognition network using user IDs. The BC condition head for DAL training uses body movement labels. Learned embeddings are also employed for attack detection.}
    \label{fig:7_bcsv_model}
\end{figure}

\smallskip
\noindent{\textbf{Parameter Selection.}} 
In Algorithm~\ref{alg:algorithm1}, there are several key parameters that should be determined to achieve the best performance. First of all, the time-frequency representation can affect the overall performance. As the STFT directly performs the Fast Fourier Transform (FFT) on the signal within a moving window, the time and frequency resolution is based on the choice of window parameters and thus, different SNR on the selected frequencies. Besides, the number of representative frequencies in air-bone domains will influence a frequency to be included or excluded. The optimal values can be determined by grid search on the development dataset. Note that there are no human efforts required to label the useful air-bone signals, the development dataset can be readily collected only if the user agrees to provide his or her speech data when using the head wearable devices as Fig.~\ref{fig:3_design_overview}. 

\smallskip
\noindent{\textbf{Threshold. }} As the last step at the TCS stage, the computed score will be compared with a predetermined threshold to generate a result. Only when the score is larger than the threshold can a user pass this stage. The threshold can be determined based on the development dataset as well.

\subsection{Stage II: BC Speaker Recognition (BC-SR) \label{sec:sec:air-bone design:s2}}
If air-bone signals pass the TCS (stage I), it means the air-bone signals are generated from the speaker's same vocalization. The next step is to verify whether this speaker is the legitimate one at this stage. Since AC signals are vulnerable to various acoustic attacks, we choose BC signals, which are insulated from acoustic domain attacks, while having unique transmission channels and containing rich speech-related information~\cite{gyrophone,liu2018vocal}, to conduct the bone conduction-based speaker recognition (BC-SR). Note that the time-frequency analysis of BC signals generates image-like representations. By adopting the deep learning image recognition technique, the BC-SR can be done by learning the speaker's unique embedding vectors under the supervision of classification loss. Therefore, we use a convolutional neural network (CNN) for feature extraction and BC speaker embedding vector generation.

\smallskip
\noindent{\textbf{Feature Extraction and Augmentation.}} 
Similar to its most successful application, namely, image classification, CNN requires its input to be convoluted with filter kernels and extracts discriminative patterns in two dimensions for recognition/verification. However, the original BC input has only one temporal dimension. To address this issue, the time-frequency representation of BC signals can expand the analysis in two dimensions. Those methods include short-time Fourier transform (STFT), continuous wavelet transform (CWT), constant-Q transform (CQT), etc. Among the choices, the STFT and CQT (with their variants) are more prevalent in speech-antispoofing challenges \cite{wu2015asvspoof,wang2020asvspoof}, 
close to this study. We employ the CQT as CNN inputs for two reasons. The first reason is CQT's better resolution in low and mid-to-low frequencies than the Mel scale \cite{lidy2016cqt}. Recall that the BC signal is mostly below 2 kHz, where a fine frequency resolution will preserve discriminative details to authenticate users. Additionally, the need to detect machine-induced BC signals implies the use of CQT, because CQT serves as an antispoofing benchmark in the ASVspoof2015 challenge \cite{wu2015asvspoof}. For better illustration, a comparison of the frequency responses of STFT and CQT is given in Fig.~\ref{fig:6_cqt_stft}, where the CQT one has better frequency resolution. Furthermore, we employ data augmentation techniques to generate sufficient and meaningful training data. Specifically, the x-axis (temporal) random flip, crop, and translation are employed, shown in Fig~\ref{fig:6_bc_cqt_aug}. Since we do not wish to change the BC spectral patterns, the augmentation along the y-axis (frequency) is not employed in this paper.

\smallskip
\noindent{\textbf{BC-SR Deep Learning Model.}} The BC-SR network is adapted from ResNet \cite{he2016deep}, with multiple stages of convolution followed by an average pooling layer. To effectively extract discriminative features from BC signals for legitimate speaker verification, we need to train the BC-SR model. Here, the training of the BC-SR model is the same as that of a classification model. The model is trained using labeled and CQT-transformed BC signals from a number of users. Cross-entropy loss is adopted to update the model parameters. In order to address the challenge of limited training data, two strategies are employed to facilitate the training process. First, although AC and BC signals are different, we assume it is still sufficient for the model to learn how to extract unique spectral information under 2 kHz from clean AC signals, i.e., the AC signals without any noises or affected by any acoustic attacks. Similar to the method in \cite{gao2021voice}, BC signals can be regarded as domain-transferred AC signals. Thus, we can pretrain the BC-SR model on public AC datasets, which have many clear voice samples (e.g., Google speech dataset~\cite{warden2018speech}). Second, inspired by the success of domain-adversarial learning (DAL)~\cite{ganin2015unsupervised,tzeng2017adversarial}, We modify the network by adding two independent heads: the speaker embedding head and the BC condition head, shown in Fig.~\ref{fig:7_bcsv_model}). As the speaker embedding head is for verification purposes, the BC condition head with the gradient reversal layer is to reduce the impacts of speaker-independent variations, e.g., different body movements. Thus, the integrated loss $L$ is
\begin{equation}
\label{eq:DAL}
L\left(w_{f}, w_{s}, w_{v}\right)=L_{s}\left(w_{f}, w_{s}\right)-\lambda \cdot L_{v}\left(w_{f}, w_{v}\right),
\end{equation}
where $L_{s}$ and $L_{v}$ represent the cross-entropy loss for speaker identification and variation classification, respectively, $w_{f}$, $w_{s}$ and $w_{v}$ indicate parameters in feature extractor, speaker embedding learner and variation learner, respectively, and $\lambda$ is a hyperparameter that controls the balance of the two losses. When BC-SR training converges, $L_{s}$ will be minimized, and $L_{v}$ will be maximized in Eq.~(\ref{eq:DAL}). This implies that the model is trained to distinguish different BC speakers while neglecting the variations caused by body movements during BC signal collection. Consequently, the trained BC-SR model can effectively learn the robust user representation.

\noindent{\textbf{BC-SR Authentication.}}  In the \textit{enrollment} phase, the legitimate user uploads a few BC signal samples. These samples will be transformed into BC speaker embedding vectors. The averaged embedding vector is then stored on the cloud server as a template. In the \textit{evaluation} phase, following the same procedure, a new embedding vector is generated for evaluation. There are two modes in BC-SR: \textit{speaker verification (SV)} and \textit{speaker identification (SI)}. The former involves the process of matching the input with a predetermined user template to authenticate the user's identity, while the latter entails assigning the input to the most probable user within a predefined, closed set of individuals. Specifically, cosine similarity score (CSS) is chosen as the distance measurement in SV, and we add a softmax and classification layer for close-set SI.

\subsection{Exploited Authentication Factors}
We summarize all exploited factors in our proposed air-bone cross-domain authentication. In TCS, by scoring the air-bone signal consistency, the \textit{behavior factor} is exploited, which can verify whether the air-bone signals are from the same speaker. Then in BC-SR, the \textit{secret biometric factor} is exploited from the BC-SR, which can ascertain whether the speaking user is the enrolled legitimate user. It should be noted that all those processes do not require any extra user efforts except speaking to control the smart voice systems.

    \begin{figure}[t]
        \centering
        \includegraphics[width = 0.9\linewidth]{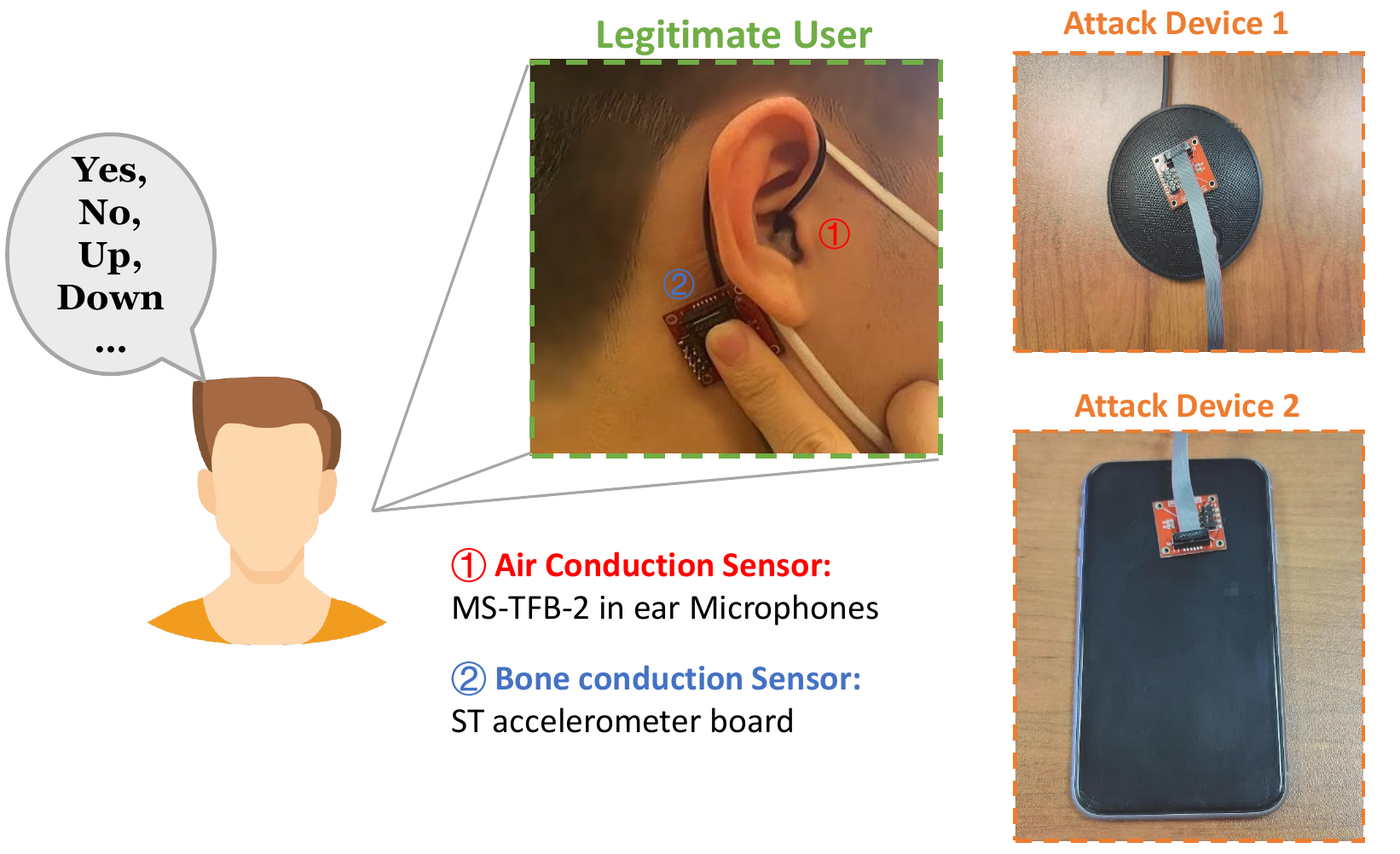}
        \caption{Air-bone data collection in the lab. We recruited 31 human subjects to collect the air-bone speaking command dataset. During data collection, every participant is instructed to put on (1) a pair of in-ear microphones to record AC signals, and (2) an accelerometer to record BC signals. Participant speaks voice commands from Google Speech Dataset \cite{warden2018speech} three times for three put-ons. In addition, we use two machine speakers to generate cross-domain attacks.}
        \label{fig:8_experiment}
    \end{figure}

\section{Performance Evaluation \label{sec:6_eval}}
\subsection{Experimental Setup}
\noindent{\textbf{Experimental Setup.}} We collected the air-bone signals via a microphone and a motion sensor. An in-ear stereo microphone MS-TFB-2 was utilized for voice recording, which is the exact position of earbuds. As for the motion sensor, we leveraged the STEVAL-MKI211V1K evaluation board with bone conduction sensor LIS25BA embedded, as well as the STEVAL-MKI109V3 motherboard for power supply and communication to a laptop. Most data processing was carried out on a laptop, while the machine learning algorithm was trained and evaluated on an RTX 8000 server. In order to implement the targeted attack scenario, we used the recorded AC voice samples to simulate the acoustic domain attacks. Specifically, a spoofed AC sample is simulated by adding the same/different user's irrelevant voice commands to the genuine one, as an acoustic replay/impersonation attack. For cross-domain attacks, we use Logitech conference speakers and iPhone built-in speakers (Devices 1 and 2, respectively in Fig.~\ref{fig:8_experiment}) as playback devices to generate spoofed air-bone signal pairs. We also tested the cross-domain attack using the laptop's internal speaker, which is not shown in the figure. To collect the bone conduction signals, the Unico GUI was exploited to communicate with the STEVAL-MKI109V3 motherboard. We then used MATLAB and Python for signal processing and model training/evaluation, respectively.

\smallskip
\noindent{\textbf{Ethics and Data Collection.}} Our experiments have been approved by our university's IRB. In total, $31$ participants were recruited to record their air-bone voice samples. During the data collection, each participant was asked to put on the in-ear microphone and accelerometer, and then speak critical commands in Google speech dataset \cite{warden2018speech} three times. Next, we asked the participant to put off all recording equipment and put them back on to record again. For each participant, the procedure was repeated for three times. In total, $8370$ spoken commands were collected from the $31$ participants. 

\subsection{Performance under Normal Use}\label{sec:performanceA}
We use Equal Error Rate (EER) as the performance metric in the proposed AirBone authentication. The threshold is set where the False Acceptance Rate (FAR), the probability that the system incorrectly accepts an unauthorized user, is equal to the False Rejection Rate (FRR), the probability that the system rejects the legitimate user. EER balances the two types of error and indicates a fair trade-off between security and usability. The system performs well with a low EER evaluation.

\subsubsection{Stage I: TCS}
We first evaluate the performance of the TCS algorithm under normal authentication attempts and false-triggering authentication attempt scenarios. For the normal authentication attempt scenario, both AC signals (i.e., voice commands) and BC signals are from the legitimate user. For the false-triggering scenario, the authentication is falsely or incidentally triggered by the AC signals from non-legitimate speakers (i.e., AC signals from environmental noises, movie playing on TV, some random conversations in the office/at home, etc.), where the legitimate user is not speaking, i.e., no coupled and meaningful BC signals. We model the normal authentication attempt scenario using coupled AC and BC signals collected from the participants and model the false-triggering scenario using the AC signals from the participants and random BC signals.

 \begin{figure*}[h]
 \centering
        \subfigure[Normal use vs. false triggering in 4 seconds.]
        {\includegraphics[width=0.3\textwidth]{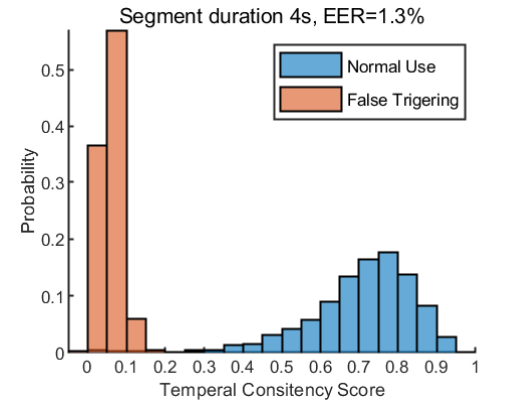}}
        \subfigure[Normal use vs. false triggering in 6 seconds.]
        {\includegraphics[width=0.3\textwidth]{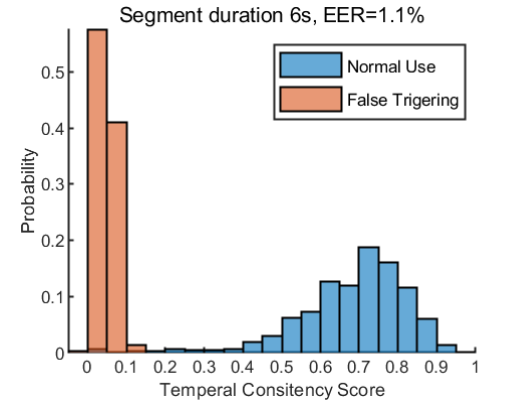}}
        \subfigure[Normal use vs. false triggering in 8 seconds.]
        {\includegraphics[width=0.3\textwidth]{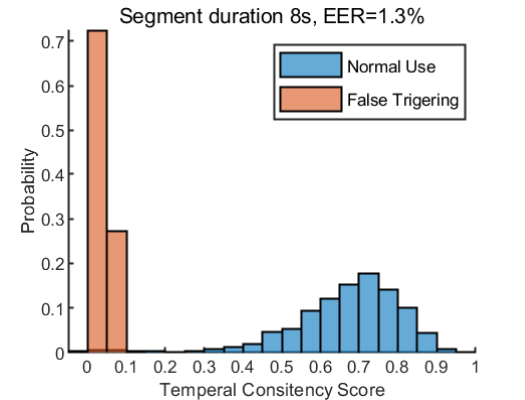}}
        \caption{Stage I TCS authentication performance under normal use situation, where the normal use is the legitimate user's air-bone signals, and the false triggering is any acoustic command (to simulate triggers invoking VA)  with random BC vibrations.}
        \label{fig:9_stage_1_eval}
\end{figure*}

 \begin{figure*}[t]
 \centering
        \subfigure[20 default/11 new users, 8s enrollment]
        {\includegraphics[width=0.3\textwidth]{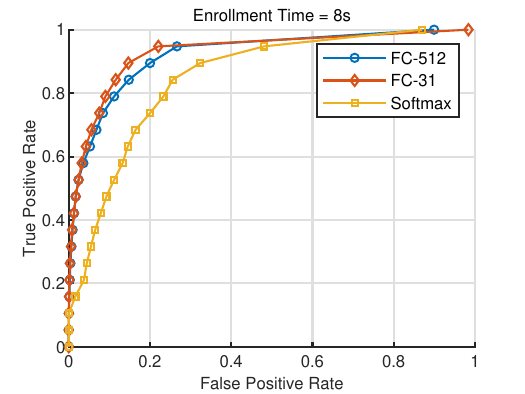}}
        \subfigure[20 default/11 new users, 24s enrollment]
        {\includegraphics[width=0.3\textwidth]{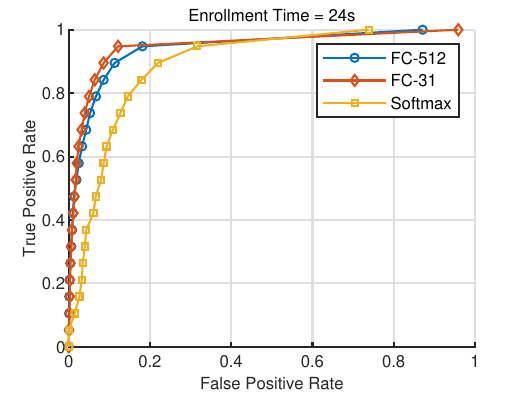}}
        \subfigure[20 default/11 new users, 40s enrollment]
        {\includegraphics[width=0.3\textwidth]{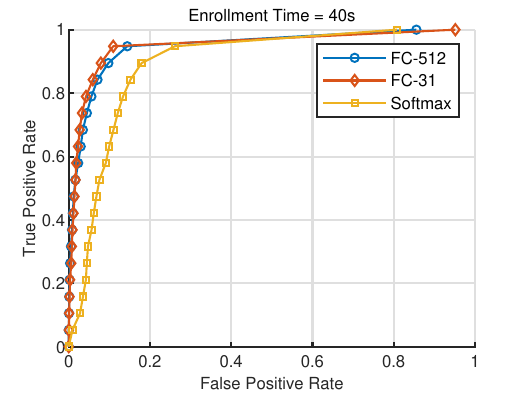}}
        \subfigure[26 default/5 new users, 8s enrollment]
        {\includegraphics[width=0.3\textwidth]{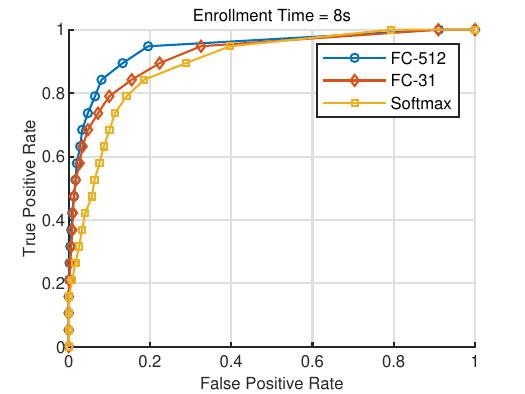}}
        \subfigure[26 default/5 new users, 24s enrollment]
        {\includegraphics[width=0.3\textwidth]{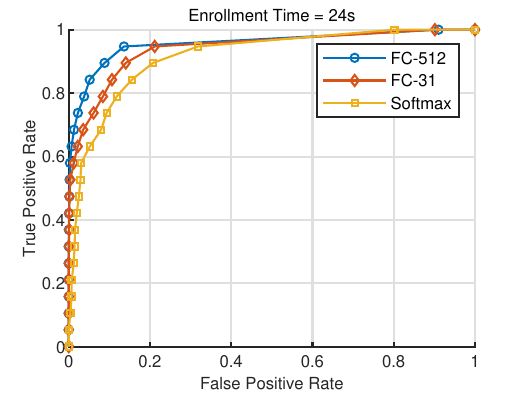}}
        \subfigure[26 default/5 new users, 40s enrollment]
        {\includegraphics[width=0.3\textwidth]{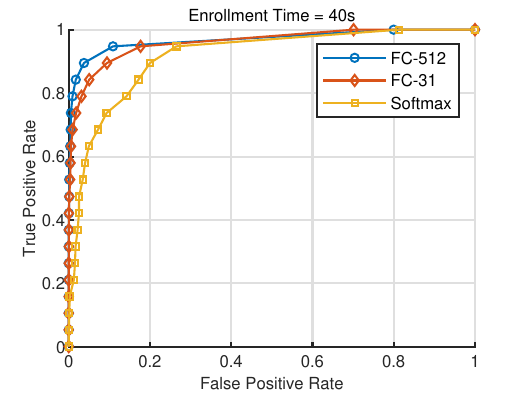}}
        \caption{Stage II BC-SR performance with a different number of default/new users and different enrollment durations. }
        \label{fig:10_stage_2_eval}
\end{figure*}

The parameters in Algorithm~\ref{alg:algorithm1} are configured to be optimal for evaluation. We set $M=N=5$ and fix the window length to be $5$ ms with an overlap of $1$ ms. The STFT is chosen over MFCC because of its computational simplicity. Next, we compute the temporal consistency score in Stage I from both normal authentication and false triggering using the whole dataset mentioned in the previous section. Besides, all air-bone data are separated into segments with a duration of 4/6/8 seconds. The histogram of the consistency scores is illustrated in Fig.~\ref{fig:9_stage_1_eval}. An interesting observation is that our proposed TCS algorithm can successfully distinguish the normal authentication attempt and false triggering with a very low equal error rate (EER) to $1.1\%$, regardless of the segment duration, even for short 4-second voice commands. Inferring from Fig.~\ref{fig:9_stage_1_eval}, we explain this observation that the proposed TCS algorithm can effectively capture representative frequencies closely correlated in air-bone domains and intend to assign high scores to this correlation. By contrast, in false triggering scenarios, non-correlated air-bone signals tend to have zero scores, which will be easily distinguished from normal use. 

\smallskip
\noindent\textbf{Impacts of Air-Bone Noises.}
We then evaluate the impacts of air-bone noises. In this experiment, the AC noise is generated from white noise, while BC noise is collected from accelerometer readings when participants are moving their heads and bodies, e.g., nodding, turning and moving. Next, we normalize the air-bone noise signals and mix them with the clean air-bone signals with certain signal-to-noise ratios (SNR) to create noisy inputs. The Stage I EER results on these noisy inputs are summarized in Table.~\ref{tab:noisy air-bone}. We found the proposed TCS algorithm is robust against AC noises even with a -10dB AC-SNR. However, when the body motion generates other non-speech bone conduction signals mentioned in Section.~\ref{sec:3_adversarial}, the EER increases to a 6\% to 7\% EER when the SNR in the BC domain is 0dB. This is because the TCS algorithm picked BC noise's representative frequencies instead of the clean BC signals. Therefore, the score from the legitimate use drops, since the air-bone signal's inconsistent components are nonnegligible in a noisy environment. One possible solution to address this issue is to leverage BC speaker-agnostic denoising technique to remove the non-speech BC noise and always focus on the right frequencies. Nevertheless, it is not common for the user to use AirBone authentication during vigorous exercise, and the system is still usable by guiding the user toward good working conditions.

\begin{table}[t]
\centering
\caption{Stage I (noisy): normal Use vs. false triggering.}
\label{tab:noisy air-bone}
\begin{tabular}{llllll}
\hline
Signal-to-Noise Ratio & (AC) clean  & 5dB & 0dB & -5dB & -10dB\\ \hline
(BC) clean & 1.4\%  & 1.4\% & 1.4\% & 1.5\% & 1.5\%   \\
5dB & 1.4\%     & 2.1\% & 2.1\% & 2.4\% & 2.3\%   \\
 0dB & 5.5\% & 5.5\%    & 5.8\% & 5.7\% & 6.2\%  \\ \hline
\end{tabular}
\end{table}

\begin{table}[t]
\centering
\caption{stage II BC-SR performance on seen/unseen users.}
\label{tab:s2 result}
\begin{threeparttable}
\resizebox{0.48\textwidth}{!}{
\begin{tabular}{llllll}
\hline
\multirow{2}{*}{Dev size} & \multirow{2}{*}{Id. Acc} & \multicolumn{3}{c}{Veri. EER} \\
 &  & (enroll)8s & 24s   &   40s \\ \hline
\multirow{3}{*}{20 user} & \multirow{3}{*}{99.7\%} & (layer-1) 13.1\% & 10.1\% & 9.0\% \\
 &  & (layer-2) \textbf{10.5\%} &\textbf{ 9.2\%} & \textbf{8.0\%} \\
 &  & (layer-3) 18.0\% & 18.2\% & 14.8\% \\ 
\multirow{3}{*}{26 user} & \multirow{3}{*}{99.3\%} & (layer-1) \textbf{11.4}\% & \textbf{9.4\%} & \textbf{7.0\%} \\
 &  & (layer-2) 15.6\% & 12.7\% & 9.8\% \\
 &  & (layer-3) 17.6\% & 15.5\% & 16.5\% \\ \hline
\end{tabular}}
\begin{tablenotes}
\item [1] Dev size: user's training data used in development.
\item [2] Each user's testing data used for evaluation.
\item [3] layer-1/2/3: fully-connected(512)/(last)/softmax.
\end{tablenotes}
\end{threeparttable}
\end{table}

\subsubsection{Stage II: BC-SR}
As shown in Fig.~\ref{fig:7_bcsv_model}, we adopt a 3-layer speaker embedding head, \textit{{FC-512, FC-31, Softmax}}, where \textit{FC} stands for fully-connected layer and the number stands for the layer width. Besides, the BC condition head comprises \textit{FC-128, FC-3, Softmax}, corresponding to three BC noisy conditions, i.e. turning/nodding/moving. The gradient computed from both heads will update the model shown in Eq.~(\ref{eq:DAL}). Finally, we prepare the CQT input based on 8 seconds BC signals, with 48 bins per octave under 2 kHz.

We pretrained the BC-SV model using published AC datasets, Google Speech~\cite{warden2018speech}. Next, we use our self-collected data to further train the BC-SV neural network. In this experiment, users are randomly split into two groups: default users and new users, where the default users are seen in the training process and the others are unseen by the BC-SV model. Next, we further split the seen user's data into train and test subsets for \textit{User Identification}, and split the unseen user's data into enrollment and evaluation subsets for \textit{User Verification}. The data partitions are detailed in Table.~\ref{tab:s2 result}. We then train the network using Adam optimizer \cite{kingma2015adam}. The learning rate is set as $0.001$, with $0.9$ gradient decay factor and $0.999$ squared gradient decay factor. The total epoch number is $60$, and the minibatch size is $64$. In order to alleviate overfitting, a random (horizontal) translation and reverse in the time domain are chosen as data augmentation.

\smallskip
\noindent\textbf{Result Analysis.}
First, we report the domain-adversarial learning's impact on model training. By tuning the hyperparameter in Eq.~(\ref{eq:DAL}) as $\lambda=0.1, 0.5, 1$ and $0$ (i.e., no DAL), the corresponding training accuracy becomes $99.9\%, 99.6\%, 89.0\%$ and $97.3\%$, respectively. Therefore, the DAL improved the BC-SV model by intentionally confusing the model on BC condition recognition. In other words, the model was guided to ignore the intra-user variances caused by various BC conditions. Then, both identification and verification results are reported in Table.~\ref{tab:s2 result}. In a 20/11 of seen/unseen user partition, the identification on 20 user test data achieves only $0.3\%$ misclassification rate, while the verification of newly enrolled 11 users achieves the lowest $9.0\%$ equal error rate, with $40$ seconds enrolling time and when the embedding is gathered from FC-31 in the speaker embedding head. When more users' data in our dataset is used to train the BC-SV model, the identification error slightly increases to $0.7\%$, while the lowest verification EER is achieved by FC-512 with $40$ seconds enrolling time.

The ROC curves of verification are shown in Fig.~\ref{fig:10_stage_2_eval}. First, more enrolling time (total duration of enrolling utterances) leads to better BC-SV performance. Second, the softmax layer, which is also the output layer in the speaker embedding head, performs the worst for the BC speaker embedding extraction. Third, with the model trained on data from more users, the verification performance on fully connected layers improves. However, our self-collected data is still limited to testing the generalization on a bigger population. We assume if tested on big data, the FC-512 layer's embedding vector will outperform the one from FC-31 since the dimension of the embedding vector should not be limited by the number of training users.

\begin{figure}[t]
        \subfigure[Acoustic impersonation attack.]
        {\includegraphics[width=0.23\textwidth]{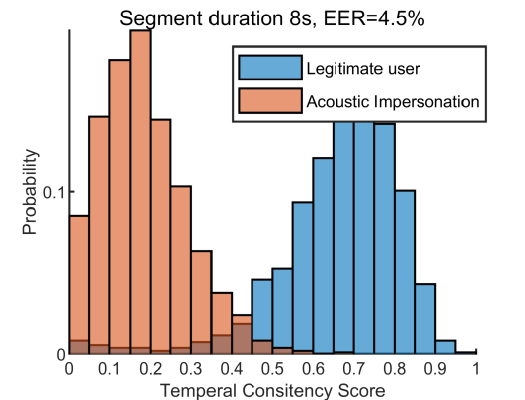}}
        \subfigure[Acoustic replay attack.]
        {\includegraphics[width=0.23\textwidth]{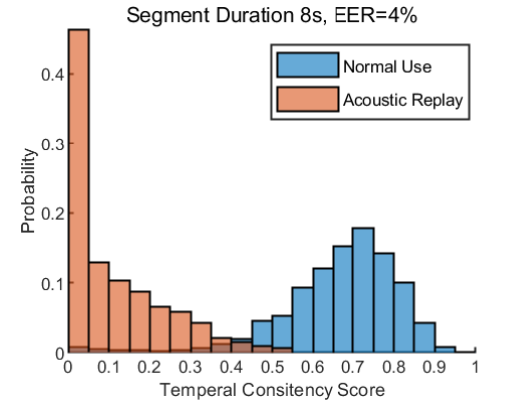}}
        \caption{Security analysis under acoustic attacks,with EER $4.5\%$ for impersonation (left) and $4.0\%$ for replay (right).}
        \label{fig:11_acoustic_attacks}
\end{figure}

\subsection{Performance under Attacks}

\noindent\textbf{Defending User-Present Acoustic Attacks.} 
We develop acoustic impersonation and replay similar to that in~\cite {shi2020wearid} as representative of acoustic attacks. This is because other spoofing attacks follow the same premise: mimicking the spectral patterns. Consequently, the discussion is solely confined to impersonation and replay attacks. Regarding the audio injection attacks, it is obvious that introducing injection will disrupt the AirBone consistency in Stage I, which can be easily detected.

We generate the attack samples by playing impersonation/replay via a loudspeaker when the legitimate user is silent. Since the acoustic attack should be detected by Stage I, we assess the TCS result by plotting the score histogram under acoustic attacks in Fig.~\ref{fig:11_acoustic_attacks}. Both acoustic attacks can be detected by the Stage I TCS algorithm with EERs lower than $5\%$, which demonstrates the TCS algorithm's ability to capture the temporal inconsistency caused by malicious attacks.

\smallskip
\noindent{\textbf{Defending Cross-Domain Attack.}}
We now evaluate the system resilience against cross-domain attacks mentioned in Section.~\ref{sec:3_adversarial}. Note that the cross-domain impersonation attack will not compromise the proposed AirBone authentication, because the BC imposter will be recognized as a stranger by BC-SR. Our focus then shifts to detecting machine-induced cross-domain attacks. For example, an attacker places the victim's device close to a loudspeaker playing voice-converted/replayed speech so that the device can collect the sound and vibration. In this case, while the AC signal is well-crafted to mimic the speech from the legitimate user, the BC signal is the loudspeaker's vibration. Since this air-bone pair is from the same source, they are indeed consistent and pass the matching-based algorithms, including the proposed Stage I. 

\begin{figure}[t]
        \subfigure[Human BC.\label{fig:12_bc_human_spec}]
        {\includegraphics[width=0.31\linewidth]{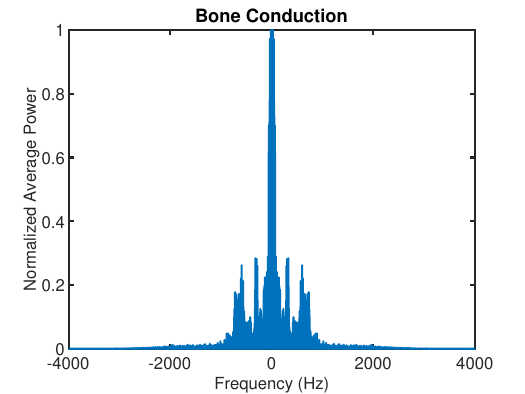}}
        \subfigure[Machine ``BC'' 1.\label{fig:12_bc_machine_spec_1}]
        {\includegraphics[width=0.31\linewidth]{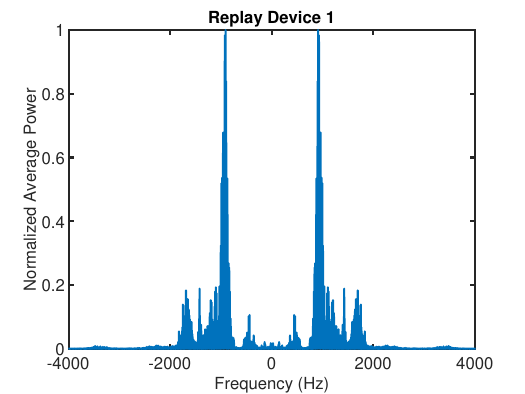}}
        \subfigure[Machine ``BC'' 2.\label{fig:12_bc_machine_spec_2}]
        {\includegraphics[width=0.31\linewidth]{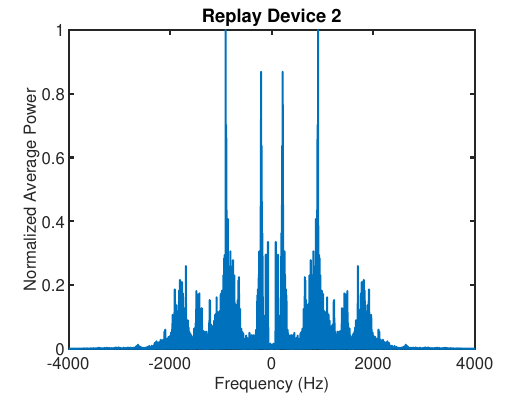}}
        \caption{Averaged spectrum of genuine BC and spoofing BC signals over 10 subjects, where the genuine BC signals are recorded directly from the human speaker, and spoofing BC signals are recorded from machines playing back recorded AC signals.}
        \label{fig:avg_spectro}
\end{figure}

We utilize the trained BC-SR model in previous evaluations and generate genuine and spoofed BC signals for machine-induced cross-domain attack detection. Specifically, the average spectrums of the real human BC signal Fig.~\ref{fig:12_bc_human_spec} and machine-induced ones are plotted in Fig.~\ref{fig:12_bc_machine_spec_1} and Fig.~\ref{fig:12_bc_machine_spec_2}. We observe that the machine-induced vibrations comprise high-frequency harmonic components that the human BC due to their different structures. Therefore, such differences can be detected by a BC-SR model trained in previous steps, which can also be reflected in the embedding vectors. To validate this, by extracting embedding vectors from vibrations collected from a Lenovo Laptop, iPhone and a Logitech conference speaker, we train classifiers including linear discriminant analysis (LDA), support vector machine (SVM), and neural network (NN) as machine-induced attack detector. Based on the results in Table.~\ref{tab:machine_detection}, our approach 
shows high detection accuracy across different classifiers, up to $99.7\%$ using LDA when tested on the conference speaker.

\begin{table}[t]
\centering
\caption{Detect Machine-induced Cross-Domain Attacks.}
\label{tab:machine_detection}
\begin{tabular}{llll}
\hline
\textbf{Machine-Induced Vibration}         & \textbf{LDA}    & \textbf{Quadratic SVM} & \textbf{Trilayered NN} \\ \hline
Laptop Internal Speaker & $99.5$\%   & $99.0$\%    & $99.2$\%        \\ 
iPhone Internal Speaker &  $97.6$\%  &  $97.0$\% &  $98.1$\% \\
Logitech Conference Speaker &  $99.7$\%  & $98.4$\%  & $99.6$\%\\ \hline
\end{tabular}
\end{table}

\subsection{Overall Performance}
We summarize the accuracy and security analysis and report the overall performance for the most common usage scenario. We first set the threshold for Stage I TCS as 0.4, where the EER is obtained when acoustic attacks happen. Therefore, the security in Stage I can be preserved without hurting the usability significantly. With this setting, the True Acceptance Rate (TAR) and False Rejection Rate can achieve $95.5\%$ and $0\%$, respectively. Even considering the acoustic attacks happening in a noisy environment, the scores of both legitimate users and attackers will decrease, which means the security performance is not getting worse in this setting. Next, in Stage II BC-SR, if the user set is small and no strangers exist, e.g., for intelligent home assistant, the user identification (detecting human attacker) and machine-induced attack detection (detecting machine attacker) achieve up to $99.7\%$ success rate. If users are from a vast open set, e.g., logging into the web portal, the speaker verification CSS threshold is set to achieve an equal error as low as $7.0\%$. However, our experiment has limited data to train and evaluate with more advanced user verification machine learning methods \cite{wan2018generalized}. We assume the user verification result can significantly increase if a sufficiently large public BC dataset is available. Finally, we employ strict aggregation from Stage I and II and report the overall performance as $95.5\%$ accuracy shown in Table.~\ref{tab:summary}.

\begin{table*}[t]
\centering
\caption{Summary and Comparison with Peer Authentication Designs.}
\label{tab:summary}
\resizebox{\textwidth}{!}{
\begin{tabular}{lllllll}
\hline
          & \textbf{AirBone}                                              & VAuth~\cite{VAuth} & XCorr$^{1}$  & EarPrint \cite{gao2021voice}     & WearID \cite{shi2020wearid}                                                           & VoiceGesture \cite{zhang2017hearing}   \\ \hline
\textbf{Exploited Features}    & BC                                                     & BC         & BC             & In-ear Voice     & Induced Vibration & Doppler Effect \\
\textbf{Auth. Factors}      & Bio+Beh & Beh          & Beh             & Bio          & Beh                                                        & Beh      \\
\textbf{Extra User Efforts} & Low                                                           & Low               & Low             & Low                      & Medium/High                                                           & Medium/High           \\
\textbf{Sensor}      & Acc                                                           & Acc               & Acc             & In-ear Mic         & Acc                                                              & Speaker+Mic    \\
\textbf{Sucess Rate}    &                                                          $95.5\%^{2}$/$7.0\%$EER$^{3}$   & $97.0$\%               & $86.3$\% & $96.4$\%            & $97.2$\%                                                      & $96.2$\%  \\ 
\textbf{False-Tri. Att. Resist.}    &                                                          $\boldcheckmark$     & $\checkmark$               & $\checkmark$ & $\times$            & $\checkmark$                                                      & $\times$   \\
\textbf{The Same Speaker Auth.}    &                                                          $\boldcheckmark$     & $\checkmark$               & $\checkmark$ & $\times$             & $\checkmark$                                                      & $\times$   \\
\textbf{The Legit. Speaker Auth.}    &                                                         $\boldcheckmark$    & $\times$               & $\times$ & $\checkmark$            & $\times$                                                      & $\checkmark$  \\
\hline

\end{tabular}}
\begin{threeparttable}
\begin{tablenotes}
\item [1] Cross-correlation matching tested with short utterances ($<8$ seconds). $^2$The aggregated two-stage result, stage I: attack resilient threshold, Stage II: identification mode. $^3$ Stage II: verification mode.
\end{tablenotes}
\vspace{-5mm}
\end{threeparttable}
\end{table*}

\section{Discussion \label{sec:7_discuss}}

\subsection{Power Consumption} 
The typical power consumption of LIS25BA is 4.6 mW in triaxial mode, 3 mW in monoaxial mode, and 0.2 mW in disabled mode. For authentication, the system can save energy by disabling the accelerometer until the keywords are identified. In addition, as implied in Section,~\ref{sec:air-bone design:s0}, BC recording only needs monoaxial mode. The underlying reasons include: (1) the bone conduction signal's propagation will generate readings in three axes; and (2) the device's facing angle towards the user is fixed. Thus, it is always possible to select the monoaxial signal for the highest quality, with respect to SNR, nonlinearity, etc. Note that Stage I performs well with air-bone segments longer than 4 seconds, whilst Stage II requires 8 seconds of BC data. Although a complete voice command must be recorded, the authentication only requires an additional 8-second BC data recording, amounting to 0.024J of additional energy consumption, which is negligible compared to the battery capacity of the equipment (e.g., 0.133-watt-hour for third-generation Airpod \cite{AirPods}). The energy it takes to wirelessly transmit a BC command is also equivalent to an eight-second extension of voice communication and can be reduced further with compression. Therefore, the proposed system's power consumption is suitable for a head wearable device.

\subsection{Computation and Memory Usage}
The computation and memory usage when using the suggested air-bone authentication system is the next issue for the system's consideration. Since only signal recording and wireless transmission are necessary and both are basic functions for a head wearable device, there is no significant cost on the user side. Furthermore, audio codecs and advanced wireless chips will considerably speed up this procedure. On the other hand, the authentication server—which is typically considered to have substantial processing power—takes on the majority of the computing and memory load for the authentication algorithm. While our experiments were successfully executed on an RTX 8000 server to authenticate a total of $31$ users, a larger user base necessitates a more complex machine learning model in Stage II BC-SR. Fortunately, this stage in the air-bone authentication may be implemented as Machine Learning as a Service (MLaaS) in order to run on a large scale. Considering the memory cost, the unit AC/BC file size in one second is around 400KB.

\subsection{Comparison with Peer Designs}
Next, we compare the proposed air-bone authentication with peer designs. Briefly, our air-bone approach can better handle short voice commands compared with the cross-correlation and matching-based algorithm in VAuth (XCorr in Table.~\ref{tab:summary}). Compared with EarPrint~\cite{gao2021voice} that may be falsely triggered by the normal operations of in-ear microphones (e.g., playing music or on a phone call), our air-bone approach is resistant to false-triggering attempts. Compared with WearID~\cite{shi2020wearid} and VAuth~\cite{VAuth}, our air-bone approach can authenticate that the cross-domain signals are not only from the same speaker but also from the legitimate human speaker. Compared with WearID~\cite{shi2020wearid} and VoiceGesture~\cite{zhang2017hearing}, both of which have to calibrate the positions of sensing devices and the user, our air-bone approach only requires the user to put on wearable devices and ``speak-to-use'', which is almost effortless. The overall comparison with peer designs is presented in Table~\ref{tab:summary}. Although our design does not achieve the highest success rate among all peer designs, the result is already usable for most applications. Most importantly, our scheme is the only one that can resist false triggering, authenticate the same speaker, and authenticate the legitimate speaker, simultaneously, which ensures security under targeted attacks. Also, with the ever-developing immersive user experiences brought by advanced IT technologies (e.g., metaverse), we expect that wearable devices such as earbuds, smart glasses, etc. will become indispensable parts of our daily lives in the near future, which further pave the way for our proposed air-bone authentication for smart voice control systems and smart voice services.

\section{Conclusion \label{sec:8_conclusion}}
In this paper, we propose AirBone, a cross-domain authentication design specifically for smart voice assistants on head-wearable devices, which is ``effortless-to-use''. For Stage I, the proposed TCS algorithm achieves $1.1\%$ EER in normal use against false triggering and $4.5\%$ EER under acoustic attacks. For Stage II, a BC-SR neural network is trained to achieve an above $99.0\%$ identification accuracy and a $7.0\%$ EER. With the attack-resisting threshold and strict aggregation, the overall success rate of AirBone is comparable to many peer designs. Moreover, AirBone can verify spoken voice commands from the same speaker's vocalization and identify this speaker simultaneously. Such a design makes the proposed authentication resistant to acoustic attacks but also cross-domain attacks in the user's absence.

\bibliographystyle{IEEEtran}

\begin{thebibliography}{10}
\providecommand{\url}[1]{#1}
\csname url@samestyle\endcsname
\providecommand{\newblock}{\relax}
\providecommand{\bibinfo}[2]{#2}
\providecommand{\BIBentrySTDinterwordspacing}{\spaceskip=0pt\relax}
\providecommand{\BIBentryALTinterwordstretchfactor}{4}
\providecommand{\BIBentryALTinterwordspacing}{\spaceskip=\fontdimen2\font plus
\BIBentryALTinterwordstretchfactor\fontdimen3\font minus \fontdimen4\font\relax}
\providecommand{\BIBforeignlanguage}[2]{{%
\expandafter\ifx\csname l@#1\endcsname\relax
\typeout{** WARNING: IEEEtran.bst: No hyphenation pattern has been}%
\typeout{** loaded for the language `#1'. Using the pattern for}%
\typeout{** the default language instead.}%
\else
\language=\csname l@#1\endcsname
\fi
#2}}
\providecommand{\BIBdecl}{\relax}
\BIBdecl

\bibitem{VoiceControlEverything}
Jasper, ``Jasper: Control anything with your voice,'' \url{https://jasperproject.github.io/}, 2021.

\bibitem{kamble2020advances}
M.~R. Kamble, H.~B. Sailor, H.~A. Patil, and H.~Li, ``Advances in anti-spoofing: from the perspective of asvspoof challenges,'' \emph{APSIPA Transactions on Signal and Information Processing}, vol.~9, p.~e2, 2020.

\bibitem{lau2004vulnerability}
Y.~W. Lau, M.~Wagner, and D.~Tran, ``Vulnerability of speaker verification to voice mimicking,'' in \emph{Proceedings of 2004 International Symposium on Intelligent Multimedia, Video and Speech Processing, 2004.}\hskip 1em plus 0.5em minus 0.4em\relax Hong Kong: IEEE, 2004, pp. 145--148.

\bibitem{ergunay2015vulnerability}
S.~K. Erg{\"u}nay, E.~Khoury, A.~Lazaridis, and S.~Marcel, ``On the vulnerability of speaker verification to realistic voice spoofing,'' in \emph{2015 IEEE 7th International Conference on Biometrics Theory, Applications and Systems (BTAS)}.\hskip 1em plus 0.5em minus 0.4em\relax Arlington, VA, USA: IEEE, 2015, pp. 1--6.

\bibitem{kinnunen2012vulnerability}
T.~Kinnunen, Z.-Z. Wu, K.~A. Lee, F.~Sedlak, E.~S. Chng, and H.~Li, ``Vulnerability of speaker verification systems against voice conversion spoofing attacks: The case of telephone speech,'' in \emph{2012 IEEE International Conference on Acoustics, Speech and Signal Processing (ICASSP)}.\hskip 1em plus 0.5em minus 0.4em\relax Kyoto, Japan: IEEE, 2012, pp. 4401--4404.

\bibitem{multifactor}
\BIBentryALTinterwordspacing
A.~Ometov, S.~Bezzateev, N.~Mäkitalo, S.~Andreev, T.~Mikkonen, and Y.~Koucheryavy, ``Multi-factor authentication: A survey,'' \emph{Cryptography}, vol.~2, no.~1, 2018. [Online]. Available: \url{https://www.mdpi.com/2410-387X/2/1/1}
\BIBentrySTDinterwordspacing

\bibitem{manzoor2019multi}
A.~Manzoor, M.~A. Shah, H.~A. Khattak, I.~U. Din, and M.~K. Khan, ``Multi-tier authentication schemes for fog computing: Architecture, security perspective, and challenges,'' \emph{International Journal of Communication Systems}, p. e4033, 2019.

\bibitem{DuoMobile}
D.~Security, ``Secure authentication with the duo mobile app,'' \url{https://duo.com/product/multi-factor-authentication-mfa/duo-mobile-app}.

\bibitem{VAuth}
\BIBentryALTinterwordspacing
H.~Feng, K.~Fawaz, and K.~G. Shin, ``Continuous authentication for voice assistants,'' in \emph{Proceedings of the 23rd Annual International Conference on Mobile Computing and Networking}, ser. MobiCom '17.\hskip 1em plus 0.5em minus 0.4em\relax New York, NY, USA: Association for Computing Machinery, 2017, p. 343–355. [Online]. Available: \url{https://doi.org/10.1145/3117811.3117823}
\BIBentrySTDinterwordspacing

\bibitem{shi2020wearid}
C.~Shi, Y.~Wang, Y.~Chen, N.~Saxena, and C.~Wang*, ``Wearid: Low-effort wearable-assisted authentication of voice commands via cross-domain comparison without training,'' in \emph{Annual Computer Security Applications Conference}.\hskip 1em plus 0.5em minus 0.4em\relax New York, NY, USA: Association for Computing Machinery, 2020, pp. 829--842.

\bibitem{gao2021voice}
Y.~Gao, Y.~Jin, J.~Chauhan, S.~Choi, J.~Li, and Z.~Jin, ``Voice in ear: Spoofing-resistant and passphrase-independent body sound authentication,'' \emph{Proceedings of the ACM on Interactive, Mobile, Wearable and Ubiquitous Technologies}, vol.~5, no.~1, pp. 1--25, 2021.

\bibitem{liu2018vocal}
R.~Liu, C.~Cornelius, R.~Rawassizadeh, R.~Peterson, and D.~Kotz, ``Vocal resonance: Using internal body voice for wearable authentication,'' \emph{Proceedings of the ACM on Interactive, Mobile, Wearable and Ubiquitous Technologies}, vol.~2, no.~1, pp. 1--23, 2018.

\bibitem{schneegass2016skullconduct}
S.~Schneegass, Y.~Oualil, and A.~Bulling, ``Skullconduct: Biometric user identification on eyewear computers using bone conduction through the skull,'' in \emph{Proceedings of the 2016 CHI Conference on Human Factors in Computing Systems}.\hskip 1em plus 0.5em minus 0.4em\relax New York, NY, USA: Association for Computing Machinery, 2016, pp. 1379--1384.

\bibitem{BC2000}
C.~P{\"o}rschmann, ``Influences of bone conduction and air conduction on the sound of one's own voice,'' \emph{Acta Acustica united with Acustica}, vol.~86, no.~6, pp. 1038--1045, 2000.

\bibitem{wu2015asvspoof}
Z.~Wu, T.~Kinnunen, N.~Evans, J.~Yamagishi, C.~Hanil{\c{c}}i, M.~Sahidullah, and A.~Sizov, ``Asvspoof 2015: the first automatic speaker verification spoofing and countermeasures challenge,'' in \emph{Sixteenth annual conference of the international speech communication association}, vol.~10.\hskip 1em plus 0.5em minus 0.4em\relax Dresden, Germany: ISCA, 2015, p. 3750.

\bibitem{wang2020asvspoof}
X.~Wang, J.~Yamagishi, M.~Todisco, H.~Delgado, A.~Nautsch, N.~Evans, M.~Sahidullah, V.~Vestman, T.~Kinnunen, K.~A. Lee \emph{et~al.}, ``Asvspoof 2019: A large-scale public database of synthesized, converted and replayed speech,'' \emph{Computer Speech \& Language}, vol.~64, p. 101114, 2020.

\bibitem{gong2019remasc}
Y.~Gong, J.~Yang, J.~Huber, M.~MacKnight, and C.~Poellabauer, ``Remasc: Realistic replay attack corpus for voice controlled systems,'' in \emph{Interspeech 2019, 20th Annual Conference of the International Speech Communication Association}.\hskip 1em plus 0.5em minus 0.4em\relax Graz, Austria: {ISCA}, Sep 2019, pp. 2355--2359.

\bibitem{zhang2017hearing}
\BIBentryALTinterwordspacing
L.~Zhang, S.~Tan, and J.~Yang, ``Hearing your voice is not enough: An articulatory gesture based liveness detection for voice authentication,'' in \emph{Proceedings of the 2017 ACM SIGSAC Conference on Computer and Communications Security}, ser. CCS '17.\hskip 1em plus 0.5em minus 0.4em\relax New York, NY, USA: Association for Computing Machinery, 2017, p. 57–71. [Online]. Available: \url{https://doi.org/10.1145/3133956.3133962}
\BIBentrySTDinterwordspacing

\bibitem{BC2010}
S.~Reinfeldt, P.~{\"O}stli, B.~H{\aa}kansson, and S.~Stenfelt, ``Hearing one’s own voice during phoneme vocalization—transmission by air and bone conduction,'' \emph{The Journal of the Acoustical Society of America}, vol. 128, no.~2, pp. 751--762, 2010.

\bibitem{BCEnhance}
M.~Li, I.~Cohen, and S.~Mousazadeh, ``Multisensory speech enhancement in noisy environments using bone-conducted and air-conducted microphones,'' in \emph{2014 IEEE China Summit \& International Conference on Signal and Information Processing (ChinaSIP)}.\hskip 1em plus 0.5em minus 0.4em\relax IEEE, 2014, pp. 1--5.

\bibitem{BCSurvey}
H.~S. Shin, H.-G. Kang, and T.~Fingscheidt, ``Survey of speech enhancement supported by a bone conduction microphone,'' in \emph{Speech Communication; 10. ITG Symposium}.\hskip 1em plus 0.5em minus 0.4em\relax VDE, 2012, pp. 1--4.

\bibitem{WPA3}
W.~Alliance, ``Wpa3 specification version 3.0,'' \url{https://www.wi-fi.org/file/wpa3-specification/}, 2020.

\bibitem{Bluetooth}
Bluetooth, ``Bluetooth technology website,'' \url{https://www.bluetooth.com/}, 2022.

\bibitem{kerberos}
B.~C. Neuman and T.~Ts'o, ``Kerberos: An authentication service for computer networks,'' \emph{IEEE Communications magazine}, vol.~32, no.~9, pp. 33--38, 1994.

\bibitem{TLS}
T.~Dierks and E.~Rescorla, ``The transport layer security (tls) protocol version 1.2,'' Tech. Rep., 2008.

\bibitem{gyrophone}
\BIBentryALTinterwordspacing
Y.~Michalevsky, D.~Boneh, and G.~Nakibly, ``Gyrophone: Recognizing speech from gyroscope signals,'' in \emph{23rd $\{$USENIX$\}$ Security Symposium ($\{$USENIX$\}$ Security 14)}.\hskip 1em plus 0.5em minus 0.4em\relax San Diego, CA: USENIX Association, 2014, pp. 1053--1067. [Online]. Available: \url{https://www.usenix.org/conference/usenixsecurity14/technical-sessions/presentation/michalevsky}
\BIBentrySTDinterwordspacing

\bibitem{lidy2016cqt}
T.~Lidy and A.~Schindler, ``Cqt-based convolutional neural networks for audio scene classification.'' in \emph{DCASE}, 2016, pp. 60--64.

\bibitem{he2016deep}
K.~He, X.~Zhang, S.~Ren, and J.~Sun, ``Deep residual learning for image recognition,'' in \emph{Proceedings of the IEEE conference on computer vision and pattern recognition}.\hskip 1em plus 0.5em minus 0.4em\relax Las Vegas, NV, USA: IEEE, 2016, pp. 770--778.

\bibitem{warden2018speech}
P.~Warden, ``Speech commands: A dataset for limited-vocabulary speech recognition,'' \emph{arXiv preprint arXiv:1804.03209}, 2018.

\bibitem{ganin2015unsupervised}
Y.~Ganin and V.~Lempitsky, ``Unsupervised domain adaptation by backpropagation,'' in \emph{International conference on machine learning}.\hskip 1em plus 0.5em minus 0.4em\relax Lille, France: PMLR, 2015, pp. 1180--1189.

\bibitem{tzeng2017adversarial}
E.~Tzeng, J.~Hoffman, K.~Saenko, and T.~Darrell, ``Adversarial discriminative domain adaptation,'' in \emph{Proceedings of the IEEE conference on computer vision and pattern recognition}.\hskip 1em plus 0.5em minus 0.4em\relax Honolulu, Hawaii: IEEE, 2017, pp. 7167--7176.

\bibitem{kingma2015adam}
\BIBentryALTinterwordspacing
D.~P. Kingma and J.~Ba, ``Adam: A method for stochastic optimization,'' in \emph{International Conference on Learning Representations (Poster)}, San Diego, USA, May 2015. [Online]. Available: \url{https://openreview.net/forum?id=8gmWwjFyLj}
\BIBentrySTDinterwordspacing

\bibitem{wan2018generalized}
L.~Wan, Q.~Wang, A.~Papir, and I.~L. Moreno, ``Generalized end-to-end loss for speaker verification,'' in \emph{2018 IEEE International Conference on Acoustics, Speech and Signal Processing (ICASSP)}.\hskip 1em plus 0.5em minus 0.4em\relax IEEE, 2018, pp. 4879--4883.

\bibitem{AirPods}
Apple, ``Airpods (3rd generation) - technical specification,'' \url{https://www.apple.com/airpods-3rd-generation/specs/}, 2021.

\end{thebibliography}

\end{document}